# Constraints on the Spin Evolution of Young Planetary-Mass Companions

Marta L. Bryan[1], Björn Benneke[2], Heather A. Knutson[2], Konstantin Batygin[2], Brendan P. Bowler[3]

[1]Cahill Center for Astronomy and Astrophysics, California Institute of Technology, 1200 East California Boulevard, MC 249-17, Pasadena, CA 91125, USA.

[2]Division of Geological and Planetary Sciences, California Institute of Technology, Pasadena, CA 91125, USA.

[3]McDonald Observatory and Department of Astronomy, University of Texas at Austin, Austin, TX 78712, USA.

**Surveys of young star-forming regions have discovered a growing population of planetary-mass (<13 $M_{Jup}$) companions around young stars[1]. There is an ongoing debate as to whether these companions formed like planets (that is, from the circumstellar disk)[2], or if they represent the low-mass tail of the star formation process[3]. In this study we utilize high-resolution spectroscopy to measure rotation rates of three young (2-300 Myr) planetary-mass companions and combine these measurements with published rotation rates for two additional companions[4,5] to provide a look at the spin distribution of these objects. We compare this distribution to complementary rotation rate measurements for six brown dwarfs with masses <20 $M_{Jup}$, and show that these distributions are indistinguishable. This suggests that either that these two populations formed via the same mechanism, or that processes regulating rotation rates are independent of formation mechanism. We find that rotation rates for both populations are well below their break-up velocities and do not evolve significantly during the first few hundred million years after the end of accretion. This suggests that rotation rates are set during late stages of accretion, possibly by interactions with a circumplanetary disk. This result has important implications for our understanding of the processes regulating the angular momentum evolution of young planetary-mass objects, and of the physics of gas accretion and disk coupling in the planetary-mass regime.**

Previous studies have sought to constrain the origin of planetary-mass companions around young stars by characterizing their mass and semi-major axis distributions, but this approach is limited by the relatively small size of the current sample[6]. Here we propose a different approach, in which we measure rotation rates for planetary-mass companions to probe their accretion histories and subsequent angular momentum evolution. In the absence of any braking mechanism, an actively accreting gas giant planet embedded in a circumstellar disk should spin up to rotation rates approaching break-up (that is, the maximum physically allowed) velocity. However, observations of the solar system gas giants indicate that they are rotating 3-4 times more slowly than their primordial break-up velocities. This may be due to magnetic coupling with a circumplanetary gas accretion disk, which could provide a channel for young planets to shed their angular momentum[7]. After the dispersal of the circumstellar and circumplanetary gas disks, late giant collisions or gravitational tides can further alter the rotation rates of some planets[8,9].

We currently have a much better understanding of the angular momentum evolution of stars, whose rotation rates have been well characterized by large surveys of star-forming regions[10]. Similar to the general picture for gas giant planets, stars spin up as they accrete material from a circumstellar gas disk. Unlike planets, stars have several known mechanisms for regulating this angular momentum, including interactions between the star and its gas disk and angular momentum loss via stellar winds[11]. Extending into the substellar mass regime, surveys of mid- to high-mass brown dwarfs (~30 – 80 $M_{Jup}$) have shown that these objects tend to rotate faster and spin down more gradually than stars, indicating that the processes that allow stars to shed angular momentum become less efficient in this mass range[12,13]. However, these studies are generally limited to brown dwarfs in nearby young clusters and star-forming regions, as most field brown dwarfs with measured rotation rates have poorly constrained ages and correspondingly uncertain mass constraints. As a result, only a handful of rotation rates have been measured for brown dwarfs with well-constrained masses less than 20 $M_{Jup}$[14,15,16], and there are no published studies of the rotation rate distribution and angular momentum evolution in this mass range.

Here, we use the near-infrared spectrograph NIRSPEC at the Keck II 10m telescope to measure rotational line broadening for three young planetary-mass companions with wide projected orbital separations: ROXs 42B b[17], GSC 6214-210 b[18], and VHS 1256-1257 b[19]. We also observe five isolated brown dwarfs that were chosen to have ages and spectral types comparable to those of the sample of planetary-mass companions: OPH 90[20], USco J1608-2315[21], PSO J318.5-22[22], 2M0355+1133[23], and KPNO Tau 4[24]. We reduce the data and measure rotation rates as described in the Methods section (Fig 1 and Table 1). We also search for brown dwarfs in the literature with spectral types later than M6, well-constrained ages typically less than 20 Myr, and measured rotation rates[25]. We use the published magnitudes, spectral types, distances, and ages to derive new mass estimates for both these objects and the NIRSPEC sample of low-mass brown dwarfs in a uniform manner (see Methods), rather than relying on the relatively heterogeneous approaches from the literature. We select our comparison sample of low-mass brown dwarfs using a cutoff of 20 $M_{Jup}$, which yields six objects with measured rotation rates including three from our survey (OPH 90, USco J1608-2315, PSO J318.5-22), and three from the literature (2M1207-3932, GY 141, KPNO Tau 12)[14,15,16]. Although this mass range includes some objects above 13 $M_{Jup}$, we note that 1σ uncertainties on the mass estimates for some of the planetary-mass companions approach 20 $M_{Jup}$, and this mass distribution is therefore consistent with that of our bound companion sample.

We compare rotational velocities for our sample of planetary-mass companions to those of the low-mass brown dwarfs. Because these brown dwarfs likely formed via direct fragmentation of a molecular cloud, systematic differences in the observed rotation rates between the two populations would suggest differing formation histories. For this analysis we also include published spin measurements for two additional planetary-mass companions, β Pic b and 2M1207-3932 b[4,5], for a total sample size of five planetary-mass companions and six low-mass (<20 $M_{Jup}$) brown dwarfs. We note that the bound brown dwarf companions GQ Lup B and HN Peg B also have measured rotation rates[26,27], but they were excluded from our sample because of their higher masses. For the objects in our observed sample we take the posterior distributions from the Markov chain Monte Carlo (MCMC) fits and divide by the probability distribution for sin*i,* where we used an inclination distribution uniform in cos*i*. Here *i* is the inclination of the object's rotational axis with respect to our line of sight. For the objects observed by previous surveys, we produce a Gaussian distribution for each rotation rate centered on the measured $v\sin i$ or $v_{eq}$ values, and for those with measured $v\sin i$ values divide that by the probability distribution for $\sin i$. This left us with a distribution of rotation rates for each object where we took into account the unknown inclination $i$. We then compare the resulting set of velocity distributions to models in which the rotational velocities of both populations are either drawn from a single Gaussian or from two distinct Gaussians using the Bayesian Information Criteria (BIC). The two Gaussian model BIC differs from the single Gaussian model BIC by $>10^3$, indicating the single Gaussian model is strongly preferred. We also calculate the Akaike information criterion (AIC) and find that the single Gaussian model is also strongly preferred, with $\Delta$AIC $> 10^3$. Finally, we calculate the evidence ratio of the two models, and again find that the single Gaussian model is favored by $>10^4$.

We conclude that at the level of our observations, there is no evidence for a systematic difference in the measured rotation rates between the sample of planetary-mass companions and brown dwarfs with comparable masses. This suggests that either the planetary-mass companions formed via the same mechanism as the brown dwarfs (that is, turbulent fragmentation), or that the processes that regulate spin are independent of formation mechanism at the level probed by our observations. This is consistent with a picture in which spin is regulated via interactions with the circumplanetary disk, as planetary-mass brown dwarfs should also host circumplanetary disks early in their lifetimes. However, it has been suggested that the properties of these disks might vary depending on the formation channel[28], and disks around isolated objects likely evolve differently than those embedded in a circumstellar disk. If spin is indeed regulated via interactions with a circumplanetary disk, our findings imply that both classes of objects should have broadly similar disk properties. We note that while there are other mechanisms such as planet-planet scattering, collisions, disk migration, and tides imposed by exomoons that could in theory alter the rotation rates of our bound companions, we do not expect any of these to affect the angular momentum evolution of these objects at the level measured by our observations.

We next compare the rotation rate for each object to its corresponding break-up velocity, taking into account uncertainties in the measured rotational line broadening, unknown inclination angles, estimated masses, radii, and ages for the objects in the sample (Fig 2). In the absence of any braking mechanism, we would expect actively accreting objects to spin up until they reach this critical rotation rate. The ratio of the observed rotation rate to the predicted break-up velocity therefore provides a useful measure of the relative efficiency of angular momentum loss mechanisms both during and after the end of accretion. Taking the error-weighted average over our sample, we find that the five planetary-mass companions that we observed are rotating at $0.137\pm0.058$ of their break-up velocity. Our sample of low-mass brown dwarfs has a similar average rotation rate of $0.114\pm0.046$ of their break-up velocity. If we combine both samples together, we find an average rotation rate of $0.126\pm0.036$ times the break-up velocity, suggesting that both populations have shed an appreciable fraction of the angular momentum acquired during accretion.

Previous studies of young stars and higher mass brown dwarfs indicate that there is a correlation between their rotation rates and masses, with lower mass objects rotating faster on average[29]. We next consider whether this correlation extends down into the planetary-mass regime, as has been suggested by previous studies[5,30]. As before, we include published rotation rates for β Pic b and 2M1207-3932 b, and show Jupiter and Saturn for reference (Supplementary Fig. 5). We exclude the terrestrial and ice giant solar system planets as their masses and spins are dominated by the accretion of solids rather than hydrogen and helium, and in some cases have been further altered by giant impacts and/or tidal evolution[8,9]. We find no evidence (Pearson correlation coefficient of -0.0788) for any correlation between rotation rate and mass for our sample of planetary-mass companions, brown dwarfs with masses below 20 $M_{Jup}$, and the solar system gas giants Jupiter and Saturn. This suggests that the mechanisms for shedding angular momentum are effectively independent of mass in the 1-20 $M_{Jup}$ range.

We next investigate how the observed rotation rates for our sample of planetary-mass companions and brown dwarfs evolve during the first several hundred Myr (Fig. 3). We find that the rotation rates for both populations appear to remain constant with respect to their break-up velocities for ages between 2-300 Myr. Furthermore, the rotation rates for these objects are also similar to the present-day rotation rates of Jupiter and Saturn; given the lack of an observed correlation between planet mass and rotation rate, this suggests that there is also no significant spin evolution on timescales of billions of years. This suggests that the observed angular velocities of planetary-mass objects are set very early in their evolutionary lifetimes, perhaps through exchange of angular momentum between the object and its circumplanetary gas disk[7].

Although the mechanism that mediates angular momentum transfer in planetary-mass companions is currently unknown, we use the observations presented here to estimate its efficiency. In the Methods section, we present a calculation that approximates the angular momentum evolution of a newly formed 10 $M_{Jup}$ object surrounded by a circumplanetary disk. Accounting for spin-up due to gravitational contraction and accretion of disk material, we find that the spin-down mechanism must extract angular momentum from the planetary-mass object at a characteristic rate of $dL/dt \sim 10^{27}$ kg $m^2/s^2$ during the disk-bearing epoch in order to reproduce the observed rotation rates in the sample. Understanding and modeling the physical nature of this mechanism represents an intriguing problem, worthy of future exploration.

The observations presented here provide constraints on the primordial rotation rates and angular momentum evolution of young planetary-mass companions and brown dwarfs with comparable masses. The degree of similarity between these two classes of objects suggests that irrespective of the formation mechanism, the physical processes that regulate angular momentum are likely to be the same for gas giant planets as they are for planetary-mass brown dwarfs. As a consequence, these observations lay the foundation for new theoretical investigations into the mechanisms that regulate gas accretion onto growing planetary-mass objects. Looking ahead, these results pave the way for future studies of gas giant planets using instruments on the upcoming generation of thirty-meter class telescopes such as the Giant Magellan Telescope's Near-IR Spectrometer.

**Acknowledgments.** The data presented herein were obtained at the W. M. Keck Observatory, which is operated as a scientific partnership among the California Institute of Technology, the University of California and the National Aeronautics and Space Administration. The Observatory was made possible by the generous financial support of the W. M. Keck Foundation. We acknowledge the efforts of the Keck Observatory staff. The authors wish to recognize and acknowledge the very significant cultural role and reverence that the summit of Mauna Kea has always had within the indigenous Hawaiian community. We are most fortunate to have the opportunity to conduct observations from this mountain. HAK acknowledges support from the Sloan Fellowship Program. Support for this work was provided by NASA through Hubble Fellowship grant HST-HF2-51369.001-A awarded by the Space Telescope Science Institute, which is operated by the Association of Universities for Research in Astronomy, Inc., for NASA, under contract NAS5-26555.

**Author Contributions.** M. L. B. led the observational program, analyzed the resulting data, and wrote the paper. B. B. helped to design and execute the observations and provided advice on the analysis as well as atmosphere models for each object. H. A. K. provided advice and guidance throughout the process. K. B. calculated the approximate angular momentum evolution of a newly formed 10 $M_{Jup}$ object surrounded by a circumplanetary disk. B. P. B. helped to identify and characterize suitable targets, including calculating new mass estimates for all of the brown dwarfs included in this study.

**Author Information.** Reprints and permissions information is available at www.nature.com/reprints. The authors have no competing financial interests to report. Correspondence and requests for materials should be addressed to mlbryan@astro.caltech.edu.

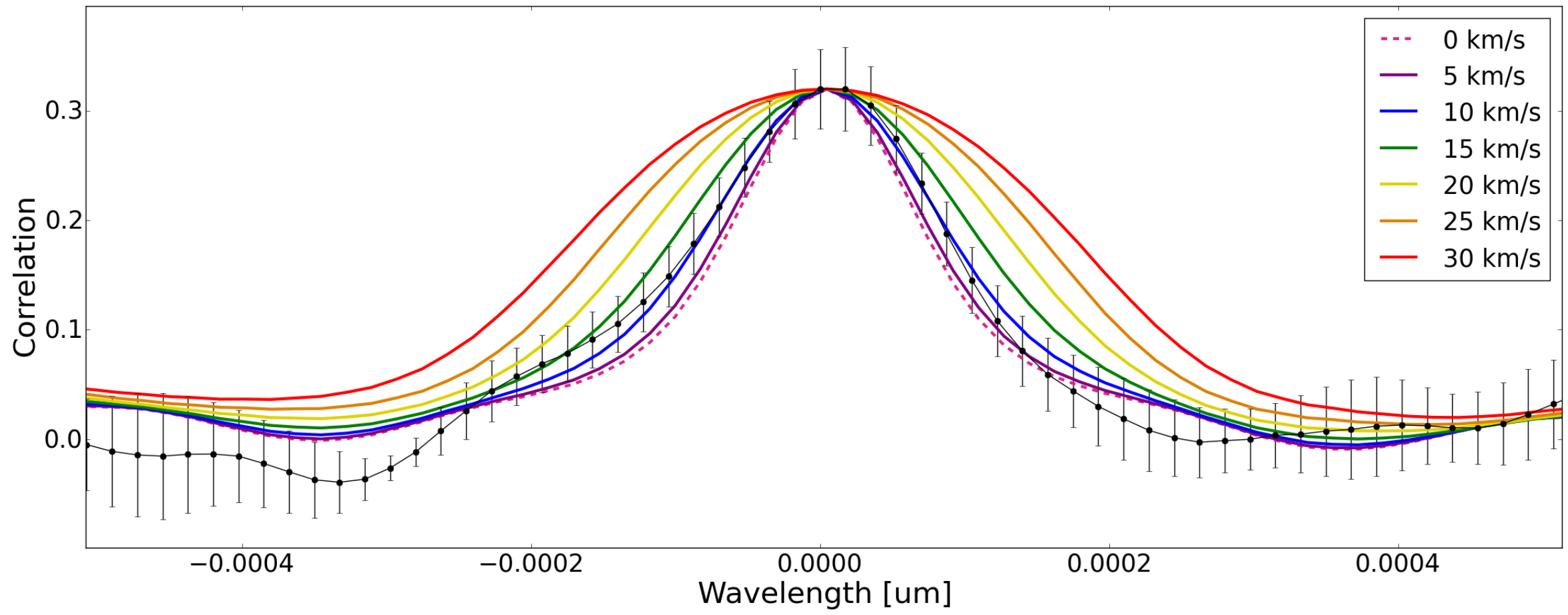

**Fig. 1. Rotational broadening in the ROXs 42B b spectrum.** Cross correlation between the ROXs 42B b spectrum and a model atmosphere broadened to the instrumental resolution (black points) with 1σ uncertainties from a jackknife resampling technique (see Methods). The cross correlation functions between a model atmosphere broadened to the instrumental resolution and that same model additionally broadened by a range of rotation rates (5, 10, 15, 20, 25, 30 km/s) are overplotted in color. The autocorrelation for a model with no rotational line broadening is shown as a dashed pink line.

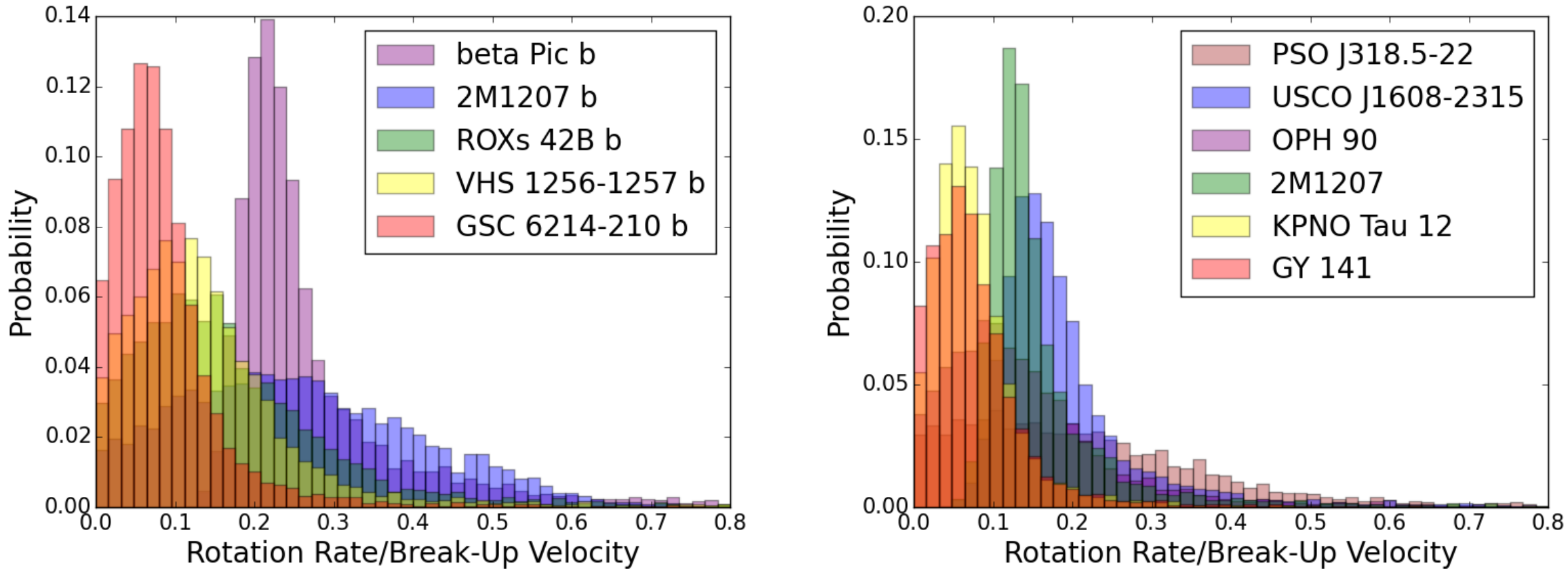

**Fig. 2. Distributions of observed rotation rates as a fraction of the corresponding break-up velocity for each object.** The distributions for the planetary-mass companions are shown in the left panel and the distributions for brown dwarfs with masses less than 20 $M_{Jup}$ are shown in the right panel. Note that these distributions take into account the uncertainties in the object's mass, age, and radius, as well as the unknown inclination of its rotation axis with respect to our line of sight. The uncertainties on the break-up velocities dominate the spread of these distributions.

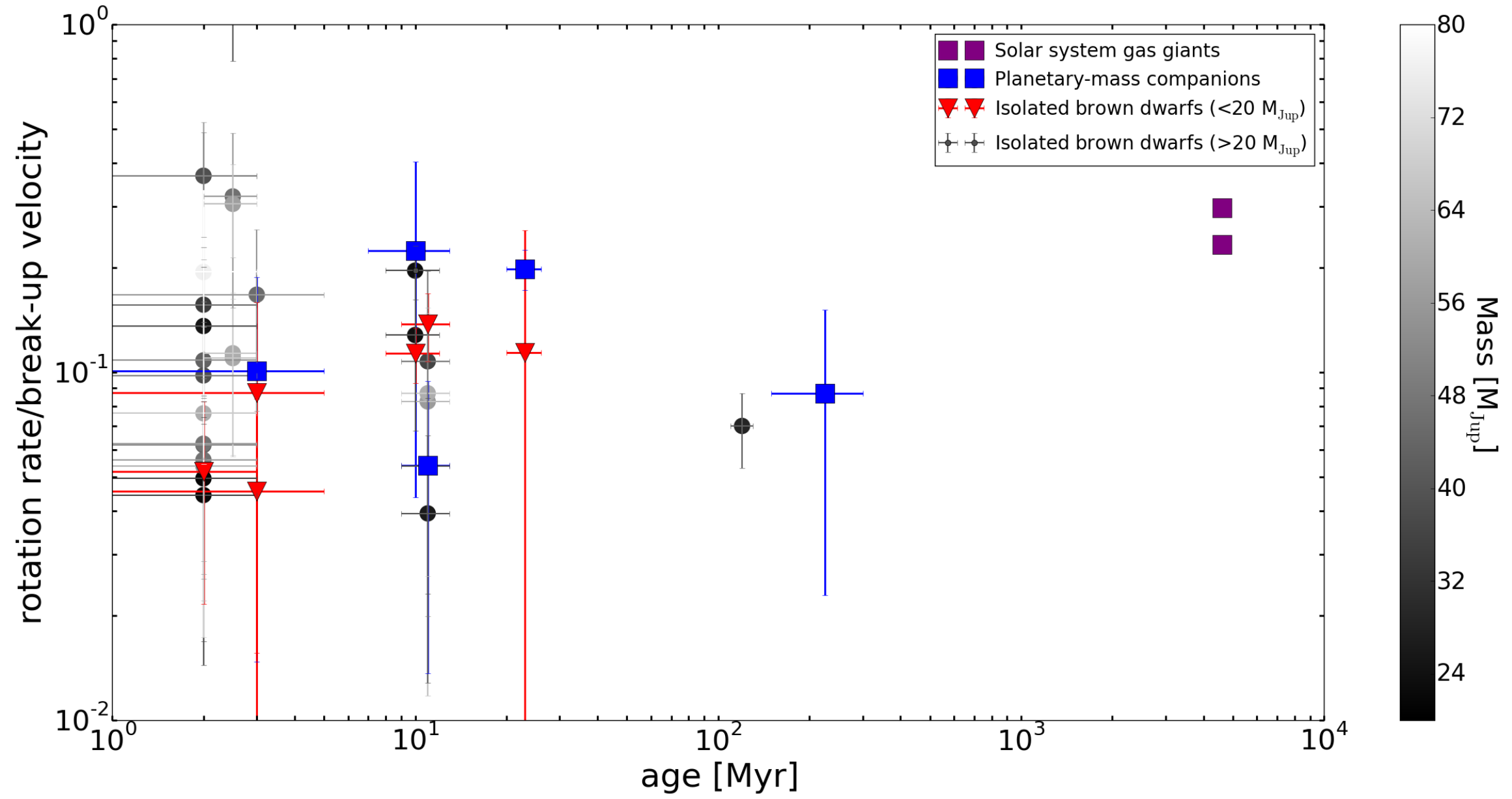

**Fig 3. Angular momentum evolution of planetary-mass objects.** Observed rotation rates as fractions of break-up velocities are plotted for our sample of five planetary-mass companions (blue squares), as well as a comparison sample of six isolated brown dwarfs with masses less than 20 $M_{Jup}$ (red triangles), and Jupiter and Saturn (purple squares). For comparison we also plot published rotation rates for all brown dwarfs with well-constrained ages typically less than 20 Myr, and spectral types later than M6 (filled circles), where the shade of grey corresponds to our new estimates of the brown dwarf masses determined using the published magnitudes, spectral types, distances, and ages of these objects. We show 1σ uncertainties for all objects; these are dominated by uncertainties in the estimated break-up velocity for each object, with an additional contribution from the measured rotation rate and unknown inclination with respect to our line of sight.

**Table 1. Measured rotation rates for our sample of three new planetary-mass companions.**

| Planetary-Mass Companion | Mass [$M_{Jup}$] | Age [Myr] | Ref. | *v*sin*i* [km/s] |
|---|---|---|---|---|
| **ROXs 42B b** | 10 +/- 4 | 3 +/- 2 | (*1*), (*17*) | 9.5 (+2.1 -2.3) |
| **GSC 6214-210 b** | 12 – 15 | 11 +/- 2 | (*1*), (*18*) | 6.1 (+4.9 -3.8) |
| **VHS 1256-1257 b** | 10 – 21 | 150 – 300 | (*1*), (*19*) | 13.5 (+3.6 -4.1) |

**Methods**

## 1. NIRSPEC Observations

We observed our targets in K band (2.03 – 2.38 um) using the near-infrared spectrograph NIRSPEC at the Keck II 10 m telescope, which has a resolution of approximately 25,000. We used the 0.041x2.26 arcsec slit for our adaptive optics (AO) observations and the 0.432x24 arcsec slit for natural seeing observations, and obtained our data with a standard ABBA nod pattern. We observed the planetary-mass companions ROXs 42B b and GSC 6214-210 b (1.2" and 2.2" separations, respectively) in AO mode in order to minimize blending with their host stars; all other targets were observed in natural seeing mode, which has a much higher (~10x greater) throughput. For ROXs 42B and VHS 1256-1257 we were able to observe both the host star and planetary-mass companion simultaneously, which made it easier to calculate a wavelength solution and telluric correction for the much fainter companions in these systems (see Methods section 2). We could not do this for GSC 6214-210 b because the planetary-mass companion was located at a separation of 2.2", which was comparable to the slit length. For this object we obtained a separate spectrum for the star after completing our observations of the companion. See Supplementary Table 1 for observation details.

## 2. 1D Wavelength Calibrated Spectrum Extraction

We extracted 1D spectra from our images using a Python pipeline modeled after ref[31]. After flat-fielding, dark subtracting, and then differencing each nodded AB pair, we stacked and aligned the set of differenced images and combined them into a single image. We then fit the spectral trace for each order with a third order polynomial in order to align the modestly curved 2D spectrum along the $x$ (dispersion) axis. For our sample of planetary-mass companions we fit the trace of the host star and used this fit to rectify the 2D spectra of both the star and the companion; this leveraged the high signal-to-noise of the stellar trace in order to provide better constraints on the shape of the fainter companion trace. Although we were not able to place GSC 6214-210 A and its companion in the slit simultaneously, we found that the shape of the spectral trace changed very little during our relatively modest 2.2" nod from the host star to the companion and therefore utilized the same approach with this data set. For both ROXs 42B b and GSC 6214-210 b, the initial solutions obtained from the stellar trace had a slope that differed by 2-3 pixels from beginning to end when applied to the companion trace. We corrected for this effect by rectifying the spectra of these two companions a second time using a linear function. Supplementary Figure 1 shows an example 2D rectified spectrum for VHS 1256-1257 A and VHS 1256-1257 b.

We note that the NIRSPEC detector occasionally exhibits a behavior, likely due to variations in the bias voltages, in which one or more sets of every eight rows will be offset by a constant value for individual quadrants located on the left side of the detector. Our GSC 6214-210 b observations were the only ones that appeared to exhibit this effect, which produced a distinctive striped pattern in the two left-hand quadrants. We corrected for this effect by calculating the median value of the unaffected rows and then adding or subtracting a constant value from the bad rows in order to match this median pixel value. While the left side of the detector remained slightly noisier than the right in our GSC 6214-210 b data set, this noise was not high enough to preclude its use in our analysis.

After producing combined, rectified 2D spectra for each order, we extracted 1D spectra in pixel space for each positive and negative trace. We calculated an empirical PSF profile along the *y* (cross-dispersion) axis of the 2D rectified order, and used this profile to combine the flux along each column to produce a 1D spectrum. For the ROXs 42B and VHS 1256-1257 datasets, which include both the star and the planet in the slit simultaneously, we plotted this empirical PSF profile and confirmed that the stellar and companion traces were well-separated in the cross-dispersion direction. We identified the range of *y* (cross-dispersion) positions containing the stellar PSF and set these to zero before extracting the companion spectrum. When extracting the host star spectrum, we similarly set the region containing the companion trace to zero.

We next calculated a wavelength solution for each spectral order. Because we maintained the same instrument configuration (filter, rotator angle, etc.) throughout the night, the wavelength solution should remain effectively constant aside from a linear offset due to differences in the placement of the target within the slit. As with the 2D traces, we leverage the increased SNR of the host star spectra to obtain a more precise solution for our sample of planetary-mass companions. We fit the positions of telluric lines in each order with a third order polynomial wavelength solution of the form: $\lambda = ax^3 + bx^2 + cx + d$, where x is pixel number. We then apply this solution to the companion spectrum using a linear offset calculated by cross-correlating the companion spectrum with a telluric model spectrum. For our brown dwarf observations we found that the SNR of the spectra was typically not high enough to obtain a reliable wavelength solution using telluric lines, and therefore determined this solution by fitting higher SNR standard star observations obtained at a similar airmass immediately before or after each observation and applying a linear offset (i.e., the same approach as for the companion spectra).

We next remove telluric lines by simultaneously fitting a telluric model and an instrumental profile to each order in the extracted spectra. For the instrumental profile, we use a Gaussian function where we allow the width to vary as a free parameter. Although we also considered an instrumental profile with a central Gaussian and four satellite Gaussians on either side[32], we found that our choice of instrumental profile had a negligible effect on our final rotational broadening measurement for ROXs 42B b and therefore elected to use the simpler single Gaussian model in our subsequent analyses. We determine the best-fit telluric models for our planetary-mass companion and low-mass brown dwarf spectra by fitting the spectrum of either the host star or the standard star, respectively, and then applying a linear offset before dividing this model from the data. Supplementary Figure 2 shows an example 1D wavelength calibrated and telluric corrected spectrum for 2M0355+1133.

**3. MCMC Fits to Determine Rotational Line Broadening**

We measure the rotational line broadening *v*sin*i*, where *v* is the rotational velocity and *i* is the unknown inclination, and radial velocity offset for each object by calculating the cross-correlation function (CCF) between the first two orders of each object's spectrum ($\lambda$ = 2.27 - 2.38 um) and a model atmosphere, where the model has first been broadened by the measured instrumental profile (R~25,000). We utilize these two orders because they contain absorption lines from both water and CO, including two strong CO bandheads, and because they have the most accurate telluric corrections and wavelength solutions. We generate atmospheric models for both our samples of planetary-mass companions and low-mass brown dwarfs using the SCARLET code[33], with the parameters used for each object listed Supplementary Table 2.

We next seek to match the shape of the measured CCF for each object by comparing it to the CCF between a model atmosphere with instrumental broadening and the same atmosphere model with both a radial velocity offset and additional rotational line broadening. We rotationally broaden the atmospheric model using a wavelength-dependent broadening kernel calculated using Equation 18.11 taken from ref.[34] for a quadratic limb darkening law. The shape of the rotationally-broadened line profile depends on the planet's limb-darkening, which varies smoothly across the covered wavelength range and between line centers and line wings. We therefore calculate limb-darkening coefficients for each of the 2048 individual wavelength bins in our spectrum using the SCARLET model. We first compute the thermal emission intensity from the planet's atmosphere across a range of different zenith angles. From those intensities we then generate model intensity profiles at each wavelength, which we fit with quadratic limb-darkening coefficients. Finally, we use the resulting limb-darkening coefficients to calculate the appropriate rotational broadening kernel at that wavelength position.

We fit for the rotational line broadening $v\sin i$ and radial velocity offset of each object using a MCMC technique. We assume uniform priors on both parameters, and calculate the log likelihood function as $\sum_{i=1}^{n}\left[-0.5\left(\frac{m_i-d_i}{\sigma_i}\right)^2\right]$, where $d$ is the CCF between the data and the model spectrum with instrumental broadening only and $m$ is the CCF of this model and the same model with additional rotational line broadening and a velocity offset applied. We calculate the uncertainties $\sigma_i$ on the CCF of the model with the data using a jackknife resampling technique:

$$(1)\quad \sigma_{jackknife}^2 = \frac{(n-1)}{n}\sum_{i=1}^{n}(x_i - x)^2$$

where $n$ is the total number of samples (defined here as the number of individual AB nod pairs), $x_i$ is the cross-correlation function calculated utilizing all but the $i$th AB nod pair, and $x$ is the cross-correlation function calculated using all AB nod pairs. The number of individual nod pairs for each target ranged between four and nine; see Supplementary Table 1 for more details.

In addition to the measurement uncertainties on our extracted spectra, we also accounted for the uncertainty on the instrumental profile in our fits to the CCF. We did this by first fitting for the instrumental profiles in individual AB nod pairs using telluric lines in our high SNR stellar spectra (either host star or standard star). We then calculated the median resolution for each night and set the corresponding uncertainty on this value to the standard deviation of all resolution values divided by the square root of the number of AB nod pairs utilized. When fitting the CCF functions of the planetary-mass companions and brown dwarfs that we observed, we drew a new resolution from a Gaussian distribution with a peak located at the median resolution and width equal to the calculated uncertainty on that resolution at each step in the MCMC chain. We plot both the measured CCFs and the best-fit model CCFs for each object in our sample in Supplementary Figure 3, and report the best-fit values and corresponding uncertainties in Table 1 and Supplementary Table 2.

We next considered whether our measured rotational broadening values might be inflated by small offsets in the relative positions of individual spectra within our sequence of AB nod pairs. We tested for this by calculating a CCF for each individual AB nod pair in our ROXs 42B b observations, where we treat the positive and negative traces separately, and measuring the location of the CCF peak in wavelength space. We found that within the set of individual positive trace spectra (A nods) and negative trace spectra (B nods) the observed wavelength shifts were minimal, typically less than 1 km/s. However, the difference between the median positive and negative trace offsets could be as large as 5 km/s. As a result we opted to fit the positive and negative trace spectra separately, resulting in two independent estimates of the rotational broadening and velocity offset for each object. We find that in all cases these two values are consistent within the errors, and report their error-weighted average in Table 1 and Supplementary Table 2. We also plot these values as a function of time in Supplementary Figure 6. As an additional check, we also used the extracted spectra for the host star ROXs 42B, which have a much higher SNR in individual exposures, to determine the spin rate for each individual AB pair. We found that our measured rotational broadening also remained consistent across the full set of AB pairs, and agreed with the value calculated directly from the composite stellar spectrum (i.e., including all AB nod pairs) to $<0.5\sigma$.

We next compare the measured radial velocity offsets for our sample of bound planetary-mass companions to those of their host stars. For the host stars, we used Phoenix spectra[35] to model their spectra, where we select the model with log(g) and $T_{eff}$ values closest to those reported in the literature for each system. We determined the rotation rates of ROXs 42B, GSC 6214-210A, and VHS 1256-1257A to be 43.6+/-0.2 km/s, 28.8+/-2.5 km/s, and 75.2 (+2.7 -2.3) km/s respectively, and measured velocity offsets of 1.8+/-0.2 km/s, -12.6 (+2.0 -2.2) km/s, and 1.5 (+2.0 -2.2) km/s respectively. We would expect both star and planetary-mass companion to share the same RV offset, as the predicted orbital velocities of these relatively wide separation companions should be much smaller than the precision of our measurements. For our lower S/N spectra (ROXs 42Bb and GSC 6214-210b), we find that the reported RV values for the planets differ from those of their host stars by 4.1 km/s and 5.3 km/s respectively. If we take the formal RV errors of 0.7 km/s and 1.3 km/s from our MCMC analysis at face value, this would correspond to RV offsets of $7.7\sigma$ and $2.2\sigma$ respectively. However, we note that for these two relatively low SNR targets, the telluric lines we use for calibrating the linear offset in the planet's wavelength solution becomes the dominant source of uncertainty in our measurement of the planet's RV offset; this is not accounted for in our formal jackknife error analysis, which assumes an error-free wavelength solution. We therefore adopt a systematic noise floor of 4.0 km/s for the reported RV values for these two relatively low SNR targets. We also test the possible effects of 4-5 km/s errors in our wavelength solutions for these two planets by setting ROXs 42Bb's radial velocity equal to that of its host star (1.8 km/s vs -2.3 km/s) and re-running our MCMC analysis. We find that the measured rotation rate for the companion in this fit is 9.8 (+2.0 -2.1) km/s, consistent with our original measurement at $0.1\sigma$.

We also investigate whether or not night-to-night variations in the instrumental broadening profile might affect our estimated values for rotational line broadening. We test this by fitting for the rotational line broadening of the host star ROXs 42B, which was observed along with its planetary-mass companion on two separate nights with an estimated instrumental resolution of R~30,000 and R~26,000, respectively. We found that the measured spins for the first and second night differed by ~1σ, indicating that our method for determining the instrumental broadening profile using telluric lines is providing a reliable characterization of this parameter.

On the modeling side, we also check whether variations in the C/O ratio of the atmospheric model used in the cross-correlation might affect our measured spin rates. We test this by repeating our CCF analysis of the ROXs 42B b spectrum using models with C/O ratios of 0.8, 0.54 (solar), and 0.35. We find that the measured spin rates for the low and high C/O models are consistent with our solar C/O model at <0.5σ. We also consider the possibility that pressure broadening might cause us to over-estimate the amount of rotational line broadening in these objects. Our fiducial solar metallicity models were generated using opacities from the ExoMol database[36], which does not include pressure broadening, but has line locations that better match our observed spectra. Alternative opacity tables such as HiTemp[37] do include pressure broadening, but do not match the line locations in our spectra as well as the ExoMol database.

We test the potential effects of pressure broadening on our estimate of the spin rate by generating a new version of our atmosphere model for ROXs 42B b using HiTemp molecular opacities and comparing the resulting $v\sin i$ value to the one measured using our original ExoMol models. Depending on our choice of pressure-temperature profile, we found that the measured rotation rate calculated using the HiTemp models was 2-5 km/s (0.7- 1.8σ) lower than the rotation rate using our original ExoMol models. We utilize the ExoMol opacities in our final rotation rate analysis for three reasons: (1) the pressure-temperature profiles for these young planetary-mass objects are poorly constrained by current observations, (2) the pressure-broadened profiles for many molecules at high temperatures are not currently well understood, and (3) the ExoMol line locations are a better match for our spectra than the HiTemp line locations. We note that including pressure-broadening in our models would likely decrease our estimated rotation rates by several km/s, corresponding to a change of approximately 1σ for most of the objects in our sample. However, this would not affect our conclusion that young planetary-mass objects appear to be rotating at much less than their break-up velocities, and that their rotation rates do not evolve significantly in time.

Finally, we test whether uncertainties in assumed effective temperatures and surface gravities could impact our measured rotation rates. We generate atmospheric models for PSO J318.5-22 using $T_{eff}$ and log(g) values determined in the forward model analysis of Allers et al 2016[43], $T_{eff}$ = 1325 (+350 -12) K and log(g) = 3.7 (+1.1 -0.1). We recalculate rotation rates and velocity offsets using a model with $T_{eff}$ = 1313 K and log(g) = 3.6, and another model with $T_{eff}$ = 1675 K and log(g) = 4.8. We find that these parameters differ from the original values by less than 0.3σ.

## 4. Calculating the Break-up Velocity

To calculate break-up velocities, we need estimates of the masses and radii of the objects in our sample. For our sample of bound planetary-mass companions we utilize mass estimates from the literature. For our sample of low-mass brown dwarfs, we derive new mass estimates in a homogeneous manner rather than relying on the heterogeneous approaches from the literature (Supplementary Table 3). We first calculate bolometric luminosities for these objects using the K-band bolometric correction for young ultracool dwarfs from ref[39] together with their distances and spectral types. We then calculate masses using their ages and luminosities together with a finely interpolated grid of hot-start evolutionary models from ref[40]. We incorporate uncertainties in distance, spectral type, apparent K-band magnitude, and age in a Monte Carlo fashion by randomly drawing these values from normal distributions for a large number of trials. We note that estimating masses using the bolometric luminosity is more robust than using absolute magnitudes as the former is less reliant on the detailed accuracy of atmospheric models. This is especially true in the optical where strong molecular opacities are generally more difficult to reproduce in synthetic spectra compared, for example, to near-infrared wavelengths.

Once we have a mass estimate, we used COND models[41] to estimate the radius of each object. We note that by using COND models, the radii we adopt assume a hot-start formation history. We calculate the 1σ minimum radius using the 1σ minimum age and mass, and the 1σ maximum radius using the 1σ maximum age and mass. Although we could have propagated the uncertainties in mass and age in quadrature, these are not independent quantities as the mass estimate depends directly on the age estimate, and we therefore opted for a more conservative approach. We calculate the best-fit break-up velocity for each object and the corresponding 1σ uncertainties on this parameter by propagating uncertainties from the mass, radius, and age of the object. Supplementary Figure 4 compares the distributions of measured rotation rates and calculated break-up velocities for each object.

## 5. Angular Momentum Evolution Calculation

Here we seek to approximately characterize the angular momentum evolution of a giant planet or low-mass brown dwarf following the primary phase of assembly. In the absence of more stringent observational constraints we utilize a simple parameterized model to estimate the relevant timescale for angular momentum evolution. It is readily apparent that this timescale will be shorter than both the disk lifetime and the Kelvin-Helmholtz timescale, but such a parameterized model is nonetheless instructive for illustrating the relevant forces at work in this problem. We begin by discussing the consequences of gravitational contraction.

### 5.1. Gravitational Contraction

To approximate the interior structure of a newly-formed planetary-mass object, we adopt a polytropic equation of state with index $\xi = 3/2$ characteristic of a fully-convective body. The binding energy of such an object is given by:

(2) $$E = -b\frac{GM^2}{R},$$

where $b = 3/(10 - 2\xi) = 3/7$, $G$ is the gravitational constant, $M$ is the mass of the object and $R$ is the radius of the object. Equating the gravitational energy loss to the radiative flux at the surface, we have:

(3) $$\frac{dE}{dt} = -4\pi R^2 \sigma T_{eff}^4 = b\frac{GM^2}{R^2}\frac{dR}{dt}.$$

Adopting an initial condition $R|_{t=0} = R_0$, the above expression yields a differential equation for the evolution of the radius:

(4) $$\frac{dR}{dt} = -R\left(\frac{R}{R_0}\right)^3\left(\frac{4\pi\sigma T_{eff}^4 R_0^3}{bGM^2}\right) = -\frac{R}{\tau_{KH}}\left(\frac{R}{R_0}\right)^3,$$

where $\tau_{KH}$ is the Kelvin-Helmholtz timescale. This equation admits the simple solution:

$$(5)\quad R = R_0 \left(\frac{\tau_{KH}}{\tau_{KH}+3t}\right)^{1/3}.$$

Using the inferred masses, radii and effective temperatures of the young (age $\leq 10$ Myr) objects within our sample, we estimate that the characteristic Kelvin-Helmholtz time for this subset of bodies is $\tau_{KH}{\sim}10$ Myr. This number has direct consequences for the spin-evolution of these objects. In isolation, conservation of spin angular momentum, $\mathcal{L}$, yields:

$$(6)\quad \frac{d\mathcal{L}}{dt} = 2IMR\omega\frac{dR}{dt} + IMR^2\frac{d\omega}{dt} = 0\ ,$$

where $I \cong 0.21$ is the reduced moment of inertia. Combined with equations (4) and (5), this expression governs the spin-up associated with gravitational contraction, and has the solution:

$$(7)\quad \omega = \omega_0 \left(\frac{3t+\tau_{KH}}{\tau_{KH}}\right)^{2/3}.$$

Importantly, expression (7) implies that in absence of external torques, the angular velocities of young planets will more than double on a timescale comparable to the Kelvin-Helmholtz time.

## 5.2. Accretion

A second process that is routinely envisioned to alter the spin evolution of young planetary mass objects is accretion of gaseous material from their circumplanetary disks. While we do not have observational constraints on circumplanetary disk lifetimes, we know that circumstellar disk dissipation timescales range from ~1 – 10 Myr. It can be reasonably speculated that circumplanetary disks exist on a similar timescale. Without observations to suggest otherwise, we assume a similar disk dissipation timescale for circumplanetary disks as seen for circumstellar disks, i.e. $\tau_{disk}{\sim}$ 3 Myr[42]. At the same time, gravitational stability requires that the ratio of the disk mass to central body's mass does not exceed the disk aspect ratio, h/r ~ 0.05. Accordingly, here we parameterize the disk mass in the following way:

$$(8)\quad M_{disk} = M\left(\frac{h}{r}\right)\exp\left(\frac{-t}{\tau_{disk}}\right).$$

Correspondingly, we interpret the derivative of the above expression as the accretion rate that the central body experiences:

$$(9)\quad \dot{M}_{disk} = \frac{M}{\tau_{disk}}\left(\frac{h}{r}\right)\exp\left(\frac{-t}{\tau_{disk}}\right).$$

In the well-studied case of circumstellar disks that encircle T-Tauri stars, stellar magnetic fields act to carve out inner gaps with a characteristic radius of ~10$R_{Sun}$[43]. While strong (~kGauss) magnetic fields and sufficient levels of ionization are essentially guaranteed in the T-Tauri setting, it is unclear whether conditions required to clear out significant magnetospheric cavities are met in typical circumplanetary disks. Accordingly, here we ignore this possibility, and assume that the disk extends down to the planetary surface, for simplicity. Under this assumption, the rate at which angular momentum is deposited upon the planet by the accretionary flow can be expressed as follows:

$$(10)\quad 2IMR\omega\frac{d\omega}{dt} = \dot{M}_{disk}\sqrt{GMR}.$$

Quantitatively, the spin-up due to gravitational contraction and that due to accretion operate on similar timescales, and have comparable magnitudes.

### 5.3. Spin-Down

While no observational constraints exist on the rotation rates of planets that are currently undergoing conglomeration, it is reasonable to anticipate that owing to accumulation of angular momentum stored in the source material, a planetary mass object should rotate at or near break-up towards the end of the phase of rapid of gas accretion (irrespective of whether the object formed through core accretion or via gravitational instability). In light of this expectation and the discussion presented above, another mechanism is needed to reduce the rotation rate to values well below break-up, and counteract spin-up due to gravitational contraction and accretion.

Because our observations do not show a statistically significant dependence of angular velocities on age, we speculate that the spin-down process (whatever it may be) operates exclusively during the disk-bearing stage of evolution. With this notion in mind, we scale this process according to the disk mass, planetary radius, and planetary spin, while parameterizing it in terms of a characteristic spin-down timescale $\tau_{spin}$:

(11) $$2IMR\omega\frac{dR}{dt} + IMR^2\frac{d\omega}{dt} = \dot{M}_{disk}\sqrt{GMR} - M_{disk}R^2\frac{\omega}{\tau_{spin}}.$$

Although this equation admits no simple analytical solution, it is readily solvable numerically.

Retaining the same parameters as those quoted above, we have solved equation (11), adopting the initial conditions $R_0 = 3\ R_{Jup}$ and $\omega_0 = \sqrt{GM/R_0^3}$ for a series of spin-down timescales in the range $\tau_{spin} = 2\times10^5 - 1.4\times10^6$ years, over a time interval of $3\tau_{disk}$. Supplementary Figure 7 shows a family of curves that denote the rotation rate of a $M = 10M_{Jup}$ planet, scaled by the breakup velocity, as a function of time. Qualitatively, these solutions exhibit the desired behavior, in that significant spin-down occurs over ~ $1\tau_{disk}$, and the spin subsequently equilibrates onto a quasi-stationary value. In particular, the evolutionary path with $\tau_{spin} = 2\times10^5$ years equilibrates onto $\frac{\omega}{\omega_{breakup}} \cong 0.1$, similar to the observed values. Accordingly, we conclude that the spin-down mechanism must operate with an approximate efficiency of $\frac{d\mathcal{L}}{dt} = -M_{disk}R^2\omega/\tau_{spin} {\sim} 10^{27}$ kg m$^2$/s$^2$ during the disk-bearing epoch, in order for sufficient spin-down to occur.

**Data Availability.** The data that support the plots within this paper and other findings of this study are available from the corresponding author upon reasonable request.

### Additional References

**Supplementary Information**

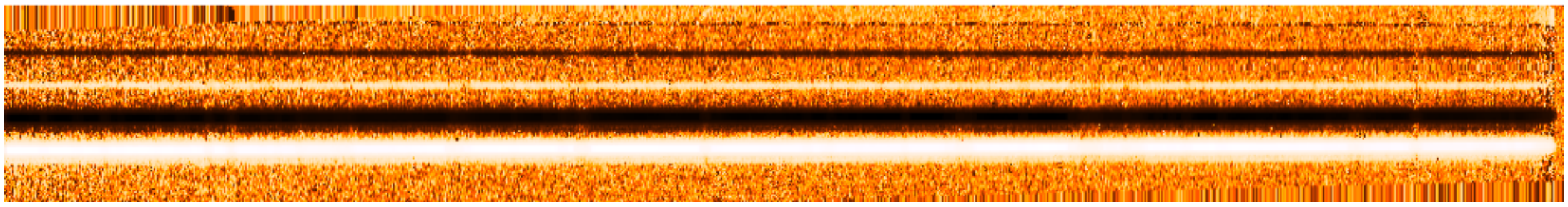

**Supplementary Figure 1. Representative 2D rectified spectrum.** 2D rectified order 1 spectrum for the system VHS 1256-1257. Both the stellar and planetary traces are visible in this spectrum.

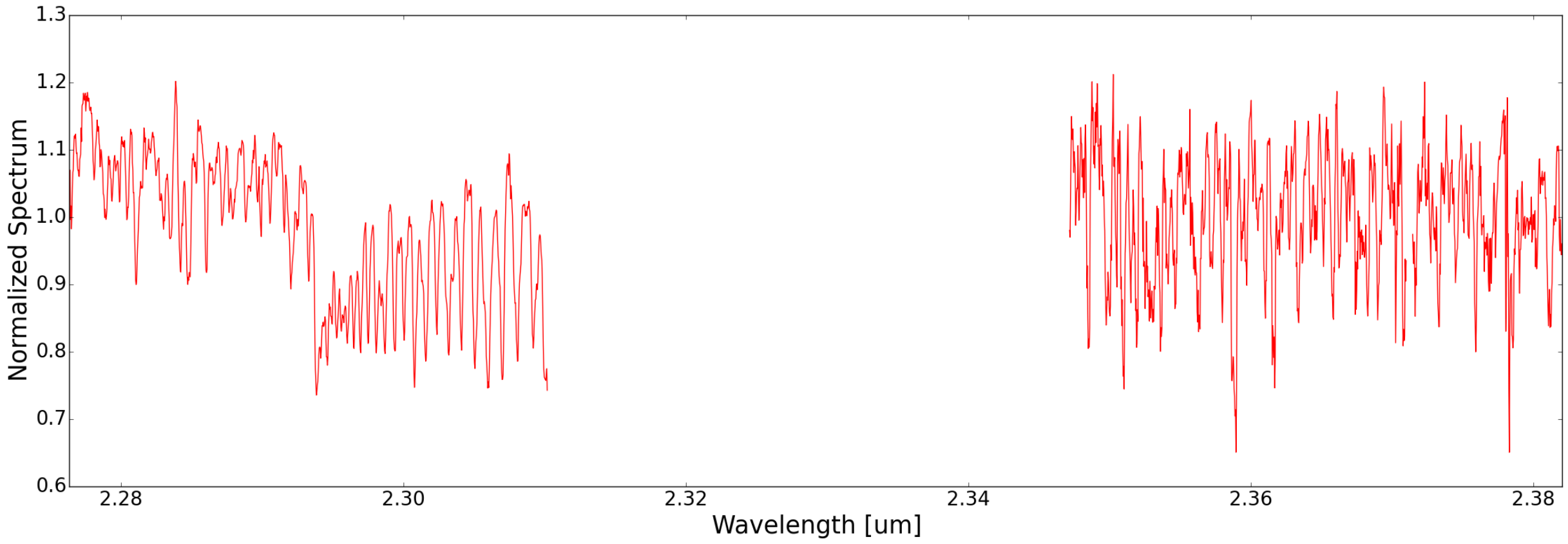

**Supplementary Figure 2. Representative wavelength calibrated and telluric corrected spectrum.** Orders 1 and 2 of the telluric-corrected spectrum for 2M0355+1133 dataset. Note the start of the CO bandhead at ~2.29 μm.

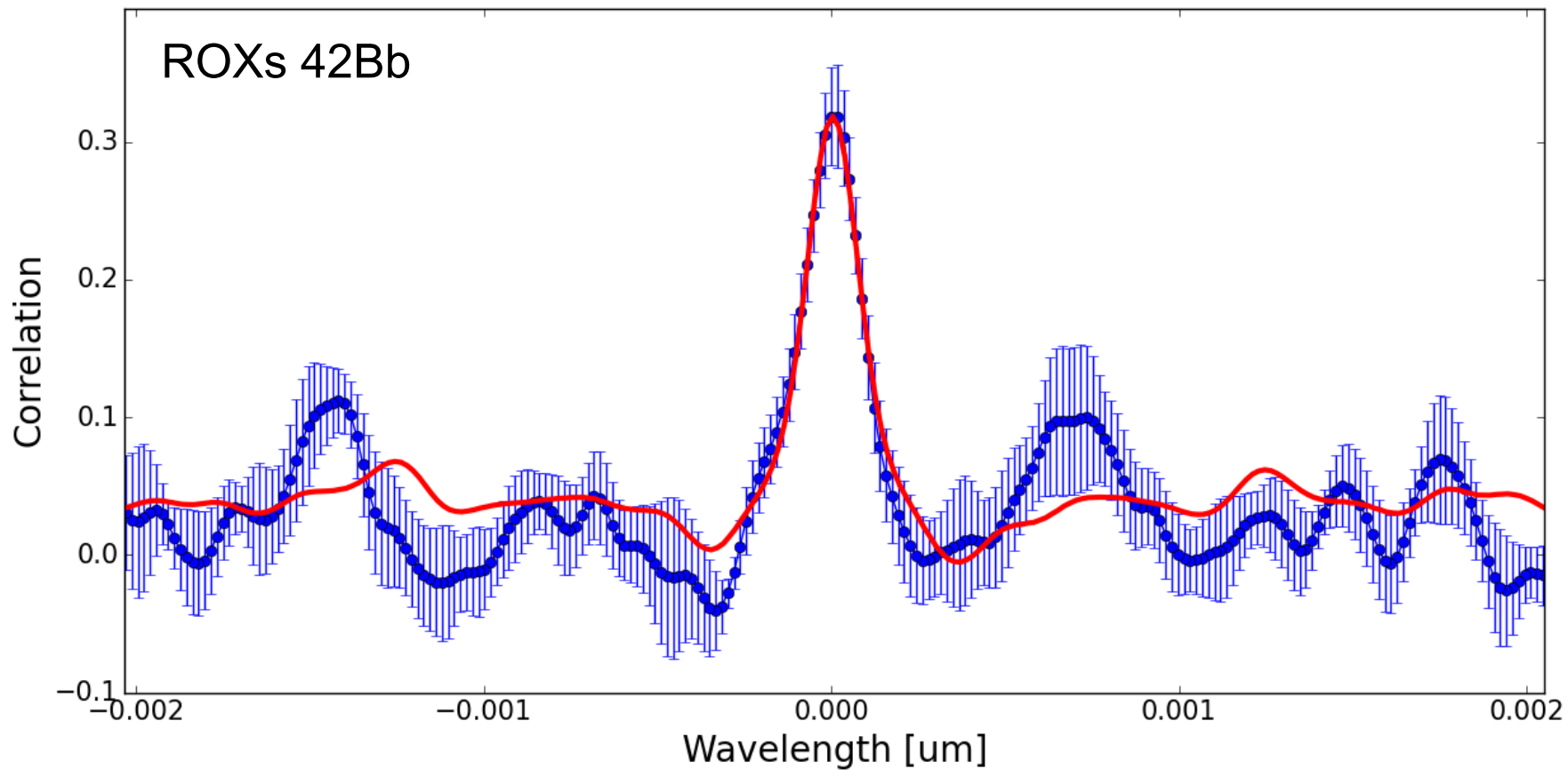

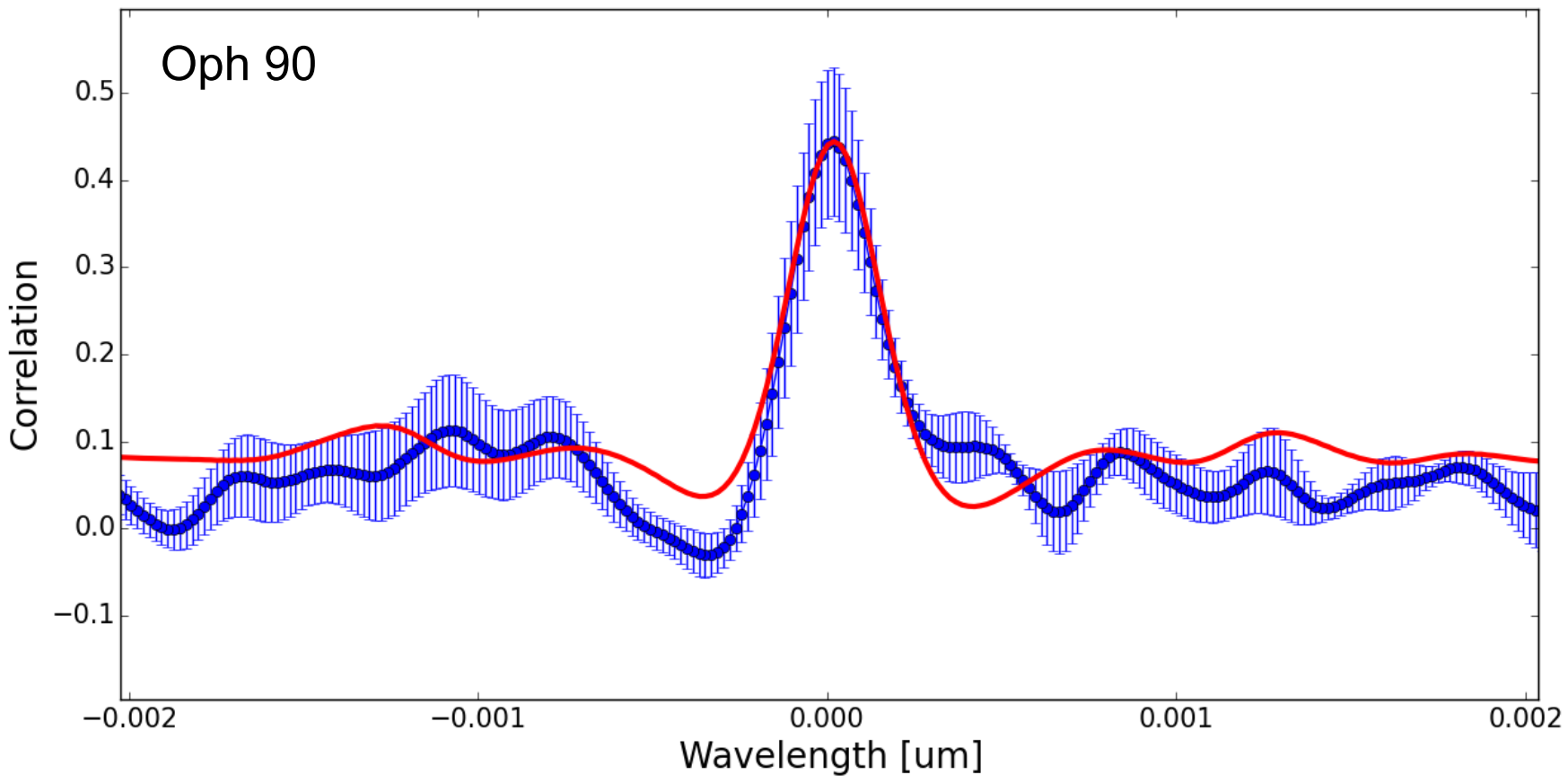

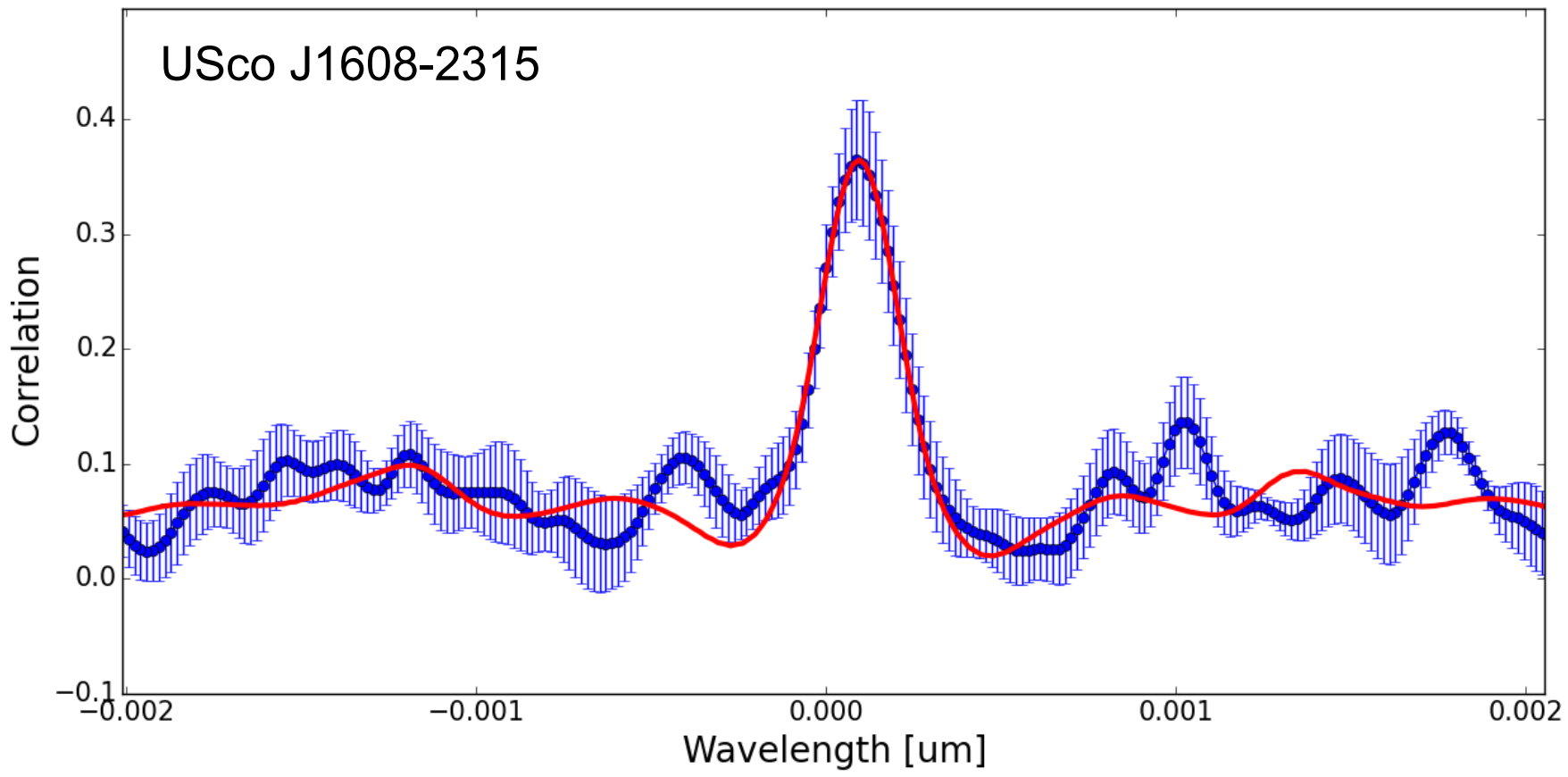

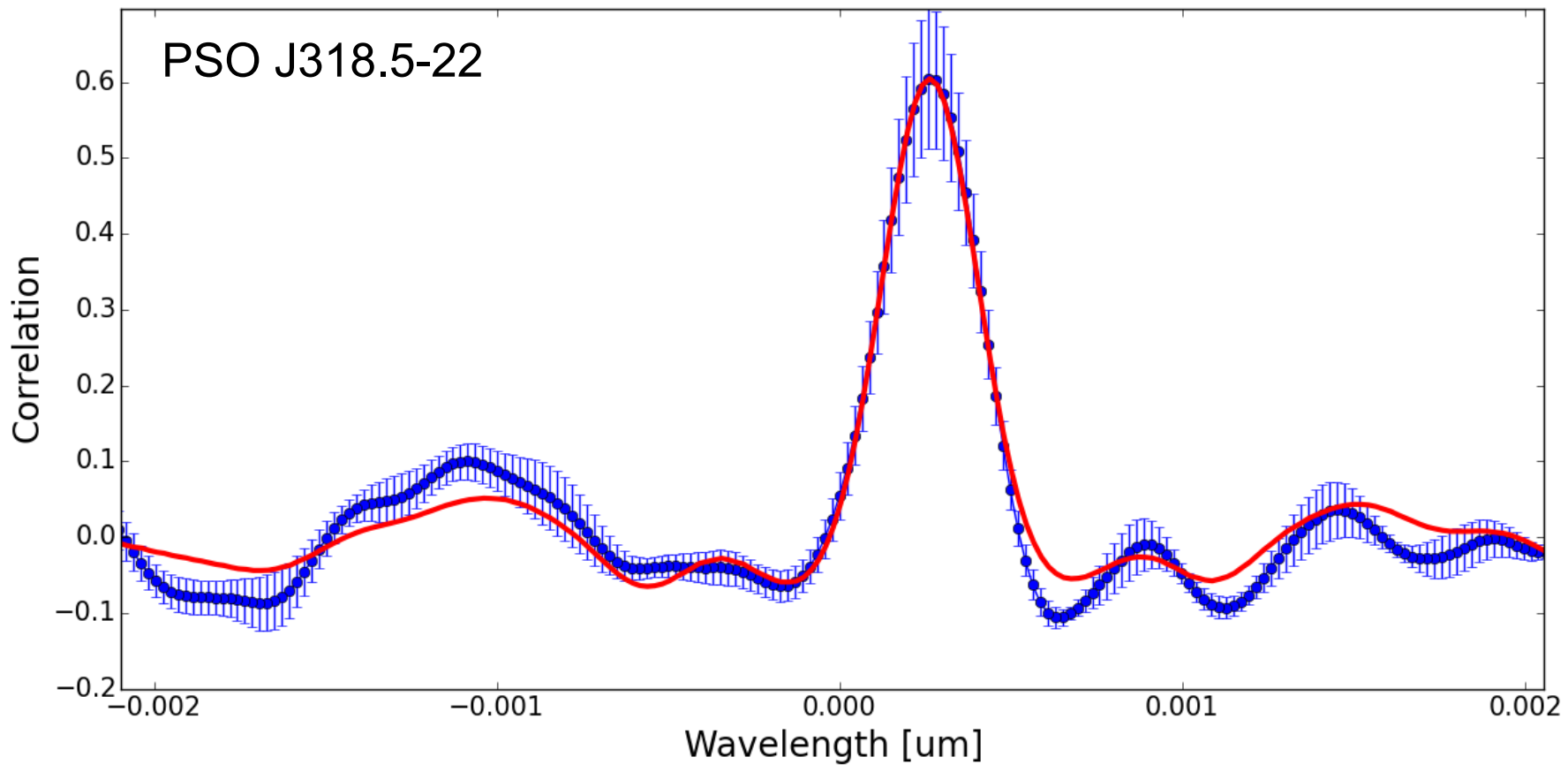

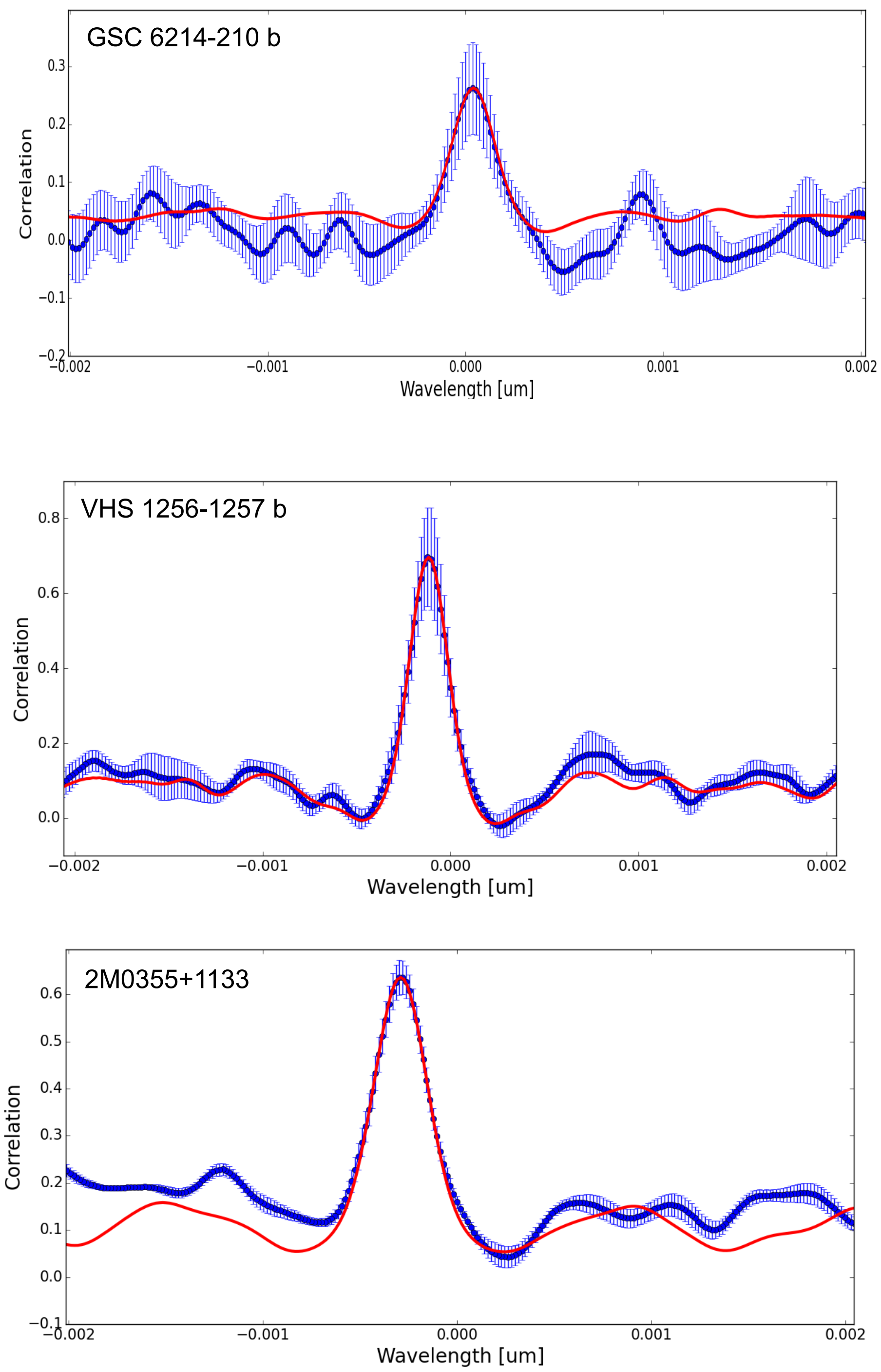

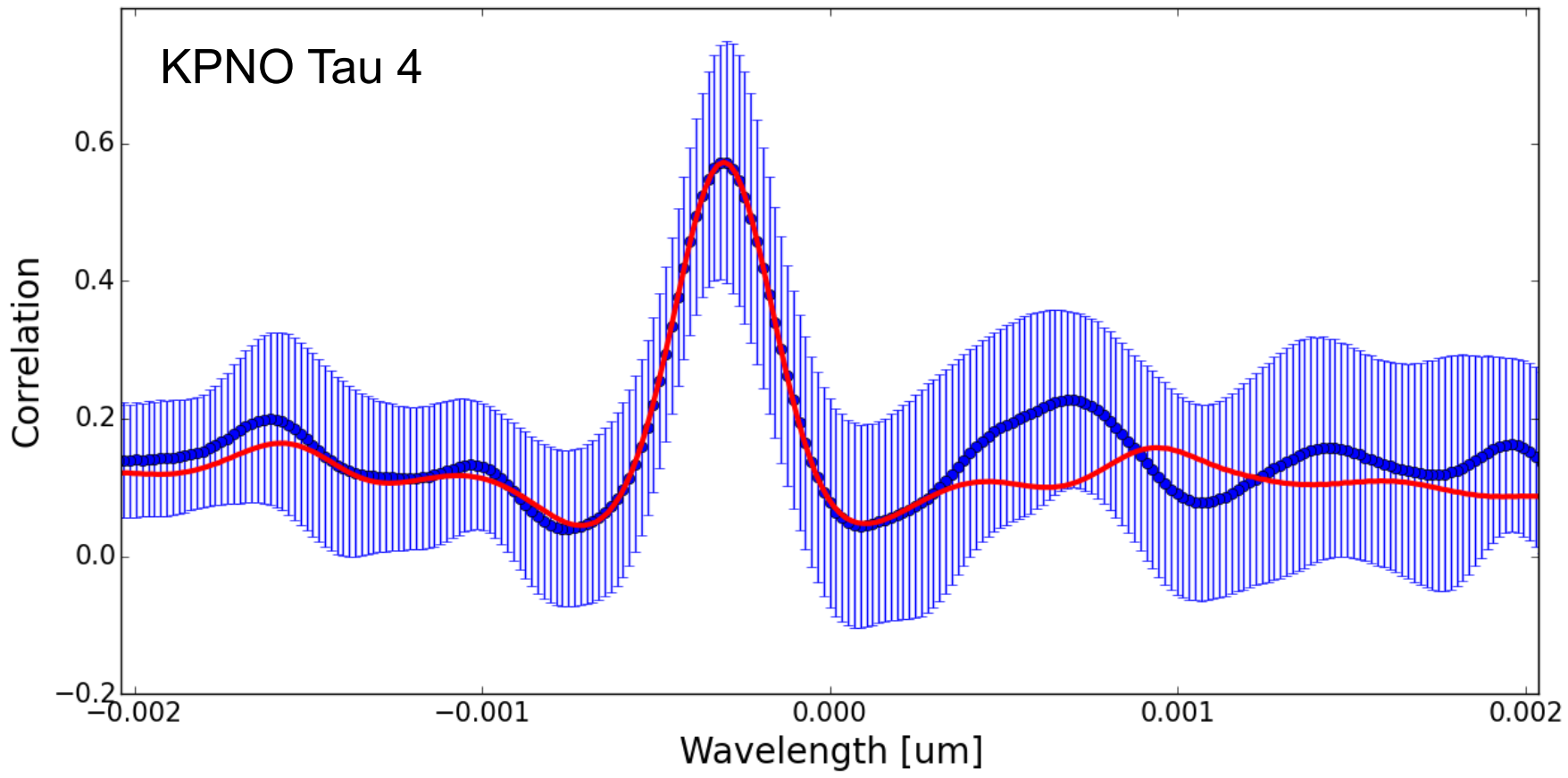

**Supplementary Figure 3. Cross correlation functions for planetary-mass companions and low-mass brown dwarfs obtained from our NIRSPEC observations.** Cross correlation functions for each object are plotted in blue, with the best-fit model overplotted in red. 1σ uncertainties on these CCFs are calculated using the jackknife resampling technique (see Methods section 3).

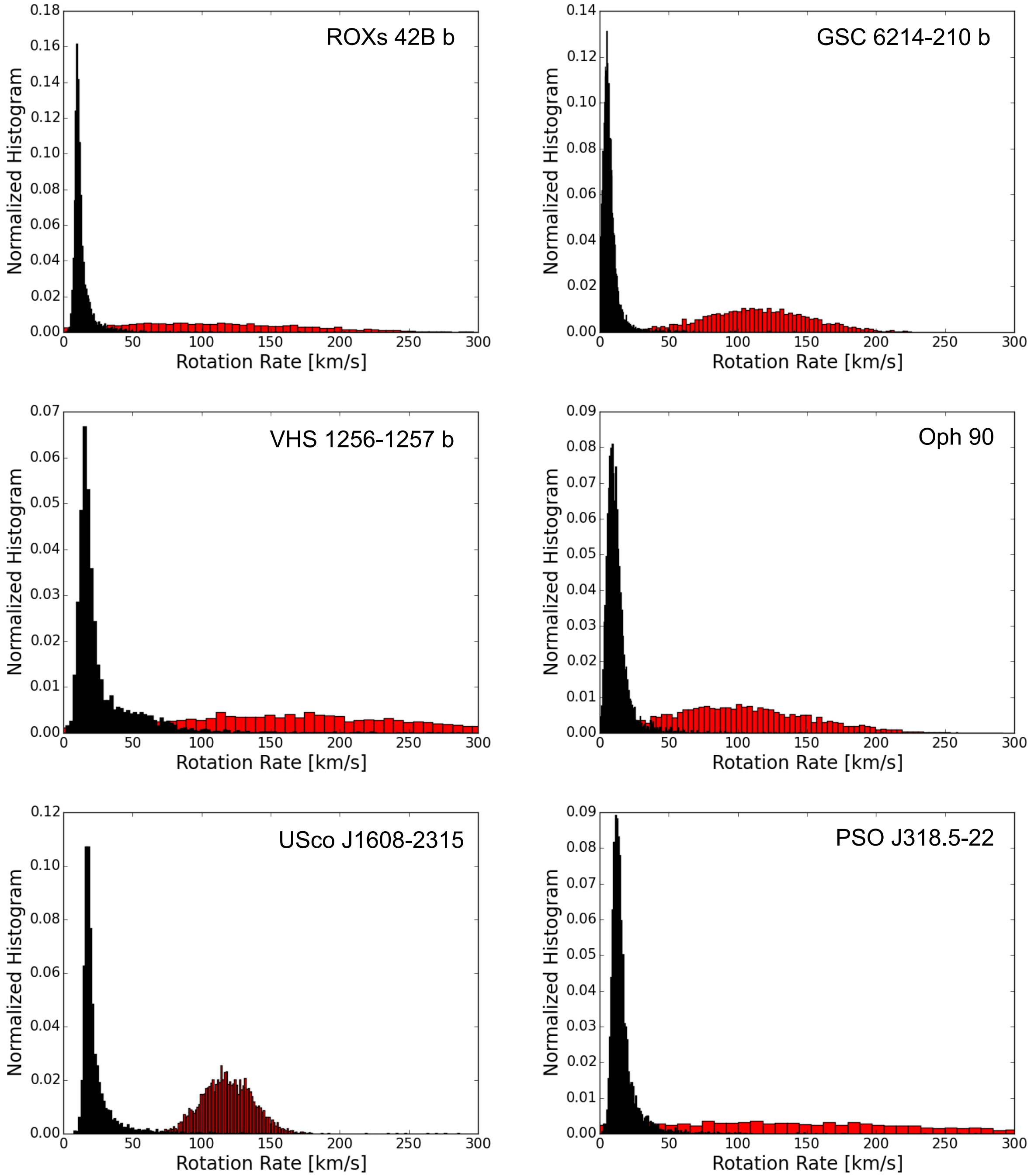

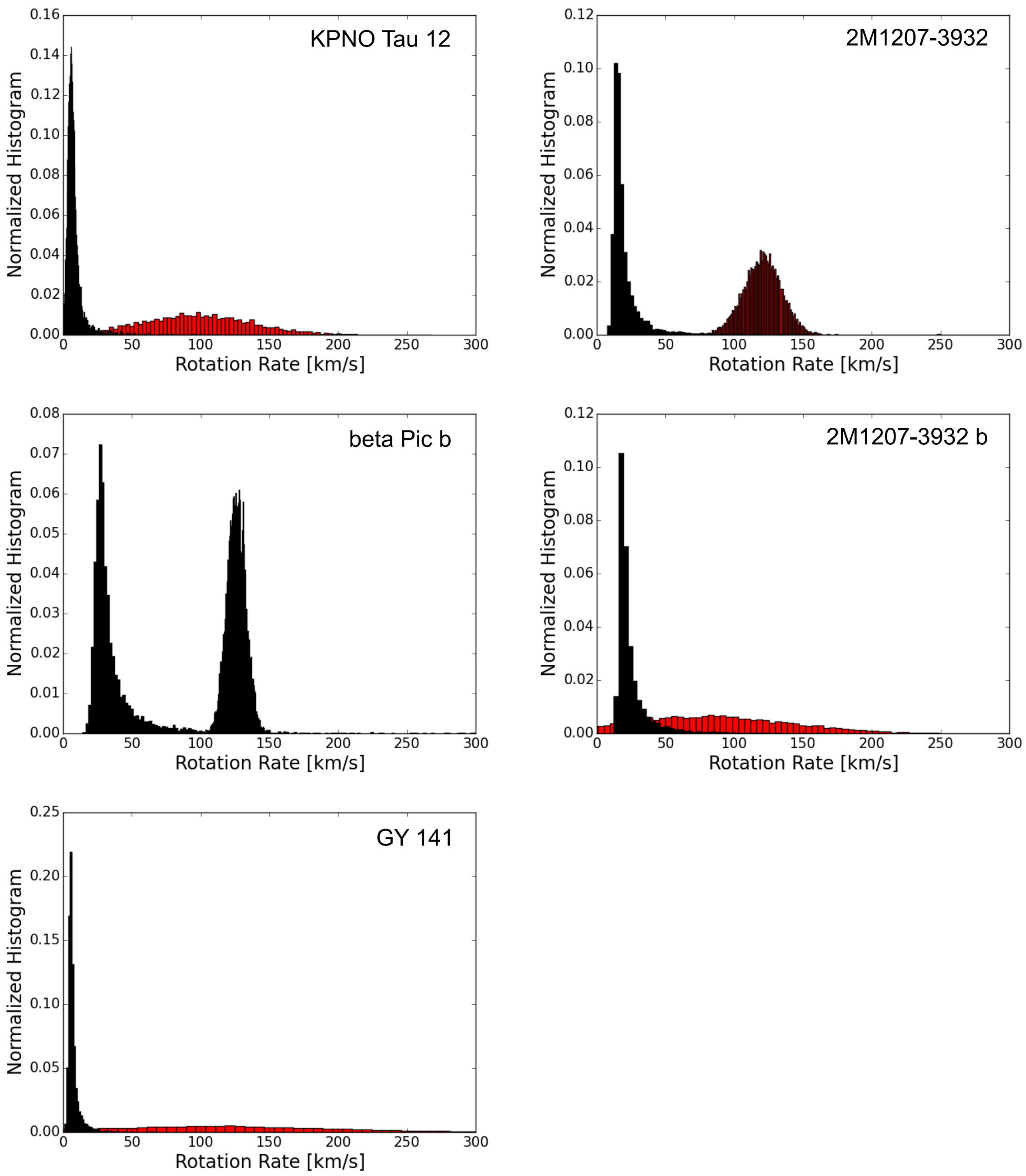

**Supplementary Figure 4. Distributions of measured rotation rates and calculated break-up velocities for each object.** The rotation rate distributions (black) have widths that are set by the uncertainties in the measured rotational line broadening and unknown inclination, and the break-up velocity distributions (red) typically have larger uncertainties that are set by errors in the estimated masses, ages, and radii.

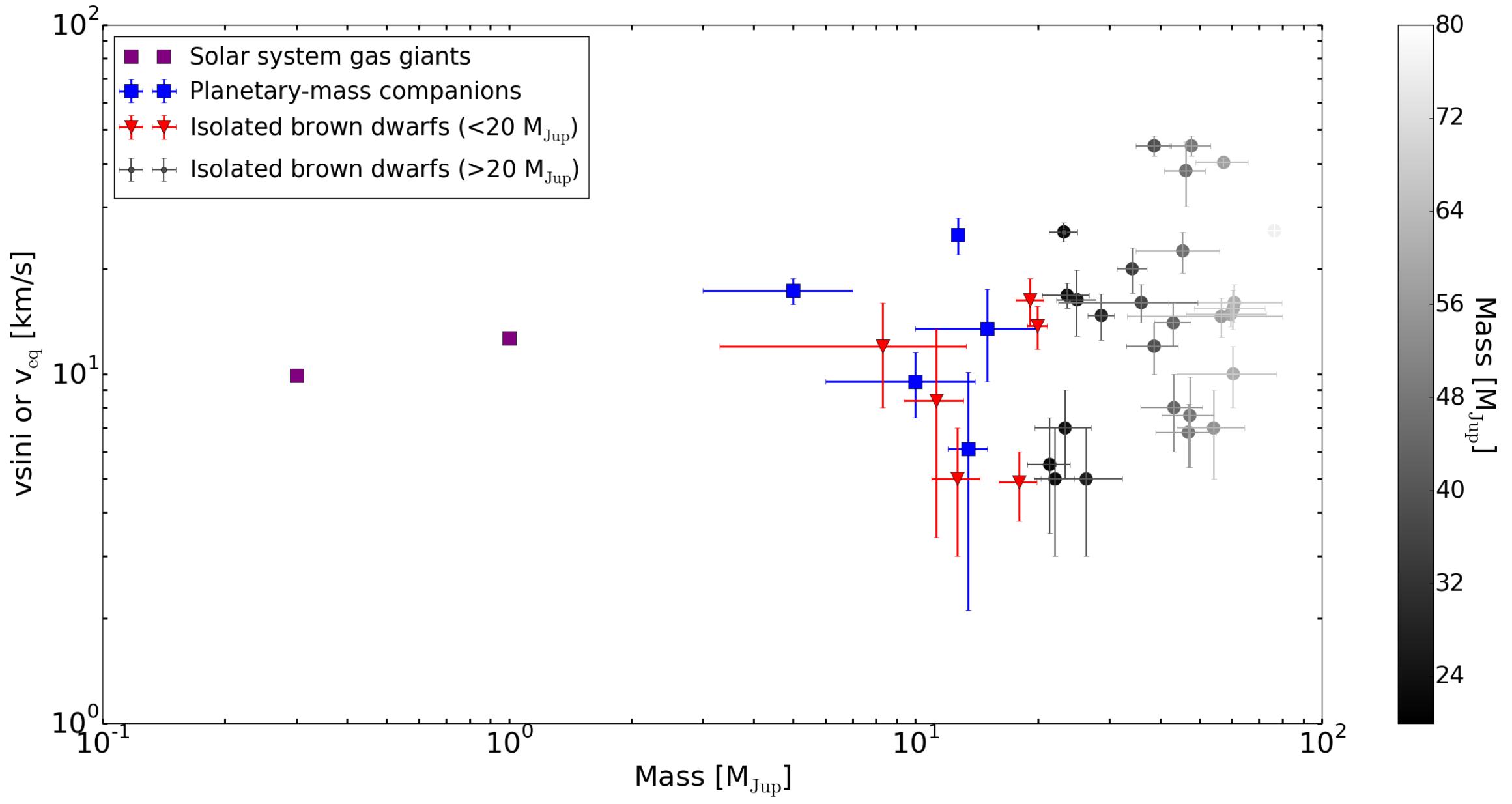

**Supplementary Figure 5. No correlation between mass and rotation rate for masses less than 20 $M_{Jup}$.** Here we show rotation rate measurements and corresponding 1σ uncertainties for the bound planetary-mass companion sample in blue and the isolated brown dwarf (<20 $M_{Jup}$) sample in red. We include the gas giant solar system planets as purple squares for reference. The rates for the brown dwarfs and all planetary-mass companions except for 2M1207-3932 b are projected velocities, and the rotation rates for 2M1207-3932 b and the solar system gas giants are equatorial velocities. We also plot rotation rate measurements for more massive brown dwarfs (20-80 $M_{Jup}$) as filled grey circles, with the shading indicating the mass of each object. Five of these measurements are equatorial velocities derived from photometric rotation periods and the rest are projected rotation rates from measurements of rotational line broadening.

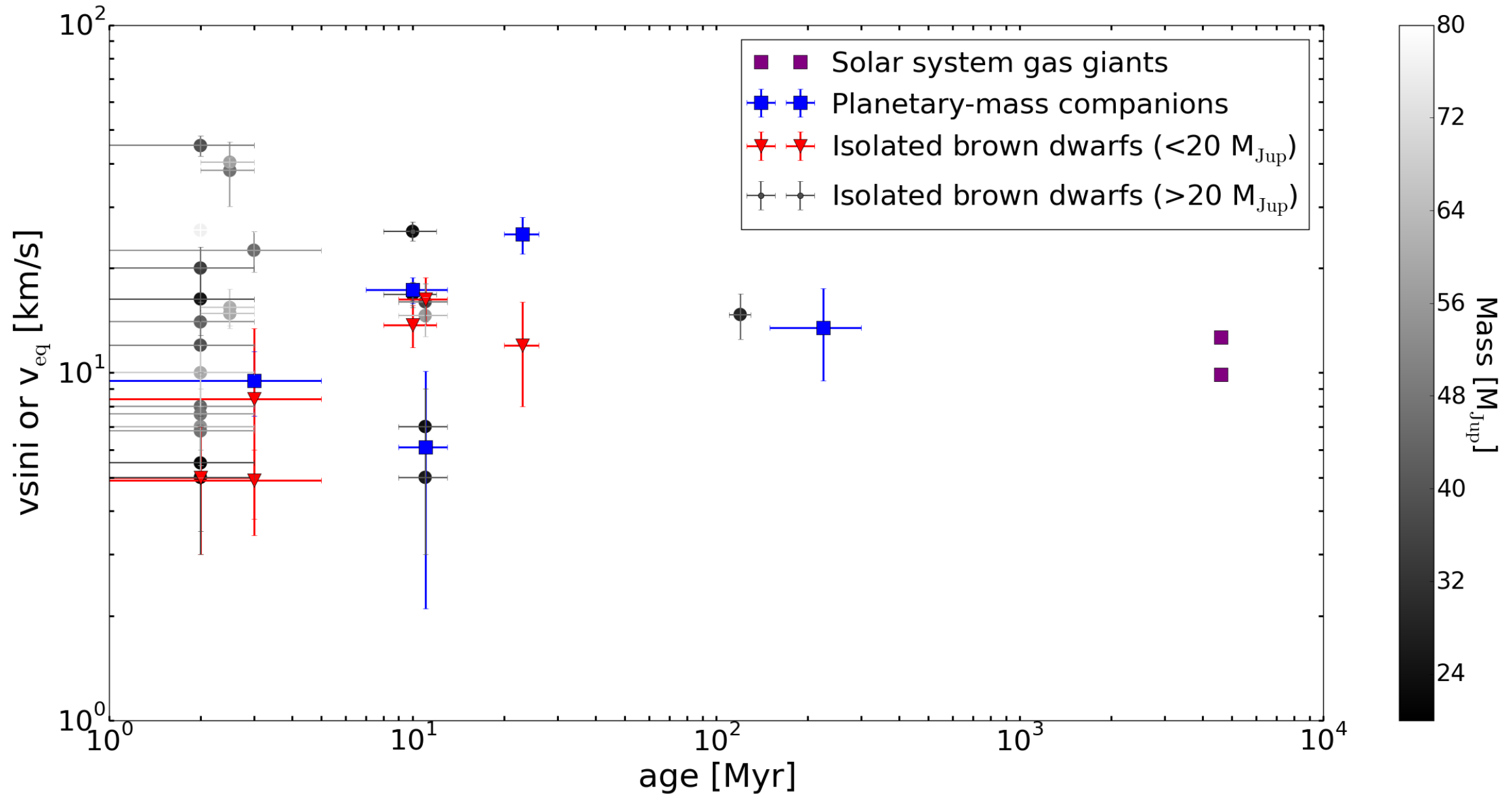

**Supplementary Figure 6.** Rotation rate measurements versus age for planetary-mass companions (blue squares) and brown dwarfs with masses less than 20 $M_{Jup}$ (red triangles). We include the gas giant solar system planets as purple squares and show more massive (20-80 $M_{Jup}$) brown dwarfs as filled circles, where color shade of grey indicates the mass. 1σ uncertainties are shown for each object.

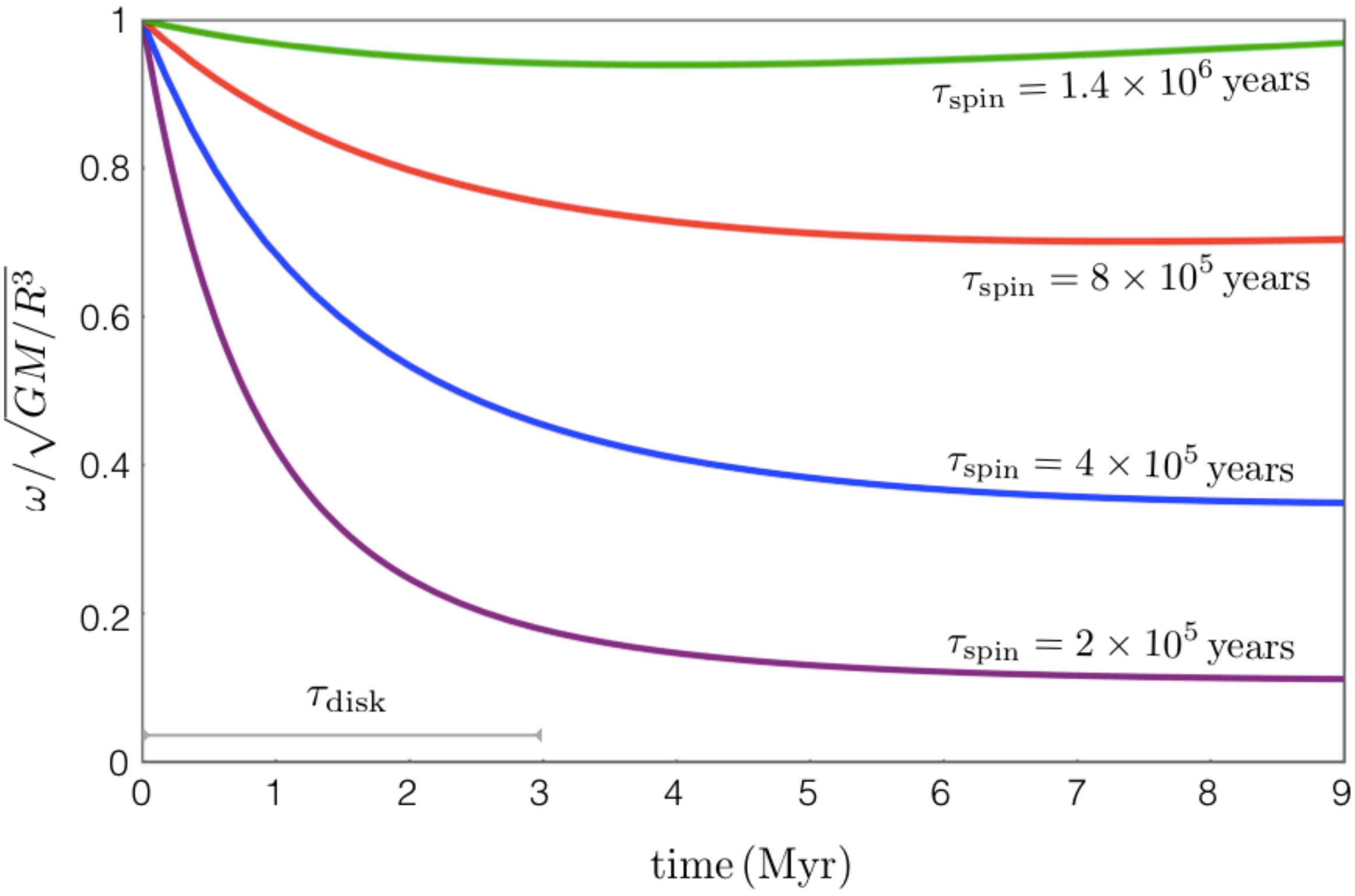

**Supplementary Figure 7. Numerical solution to equation (11).** A 10 $M_{Jup}$ planet is initialized at breakup rotation (qualitatively t=0 corresponds to the conclusion of rapid gas accretion), and subsequently experiences spin-up due to gravitational contraction and accretion, as well as spin-down due to a parameterized angular momentum exchange with the circumplanetary disk.

**Supplementary Table 1.** ***NIRSPEC K*** **Band Observations.**

| System | Pri. SpT | $m_{K,star}$ (mag) | $m_{K,pl}$ (mag) | Pl. SpT | Sep. (",AU) | $M_{comp}$ ($M_{Jup}$) | Age (Myr) | UT Date | AO? | No. Exp | Tot. Exp. (min) | Pl. S/N |
|---|---|---|---|---|---|---|---|---|---|---|---|---|
| **ROXs 42B** | M0 | 8.7 | 15.0 | L1 | 1.2,140 | 10+/-4 | 3+/-2 | 2015/6/1 | Yes | 18 | 233 | 7.4 |
| **ROXs 42B** | M0 | 8.7 | 15.0 | L1 | 1.2,140 | 10+/-4 | 3+/-2 | 2015/6/2 | Yes | 6 | 80 | 4.4 |
| **GSC 6214-210** | M1 | 9.2 | 14.4 | L1 | 2.2,320 | 12-15 | 11+/-2 | 2015/6/3 | Yes | 16 | 240 | 6.0 |
| **VHS 1256-1257** | M7.5 | <10.4 | 14.7 | L7 | 8.1,102 | 10-21 | 150-300 | 2015/5/7 | No | 14 | 93 | 11.1 |
| **OPH 90** | … | … | 14.9 | L0 | … | 11+/-2 | 3+/-2 | 2015/6/4 | No | 8 | 52 | 16.6 |
| **USco J1608-2315** | … | … | 14.2 | L1 | … | 19+/-1.5 | 11+/-2 | 2015/5/7 | No | 12 | 100 | 12.3 |
| **PSO J318.5-22** | … | … | 14.4 | L7 | … | 8.3+/-0.5 | 21+/-4 | 2015/6/4 | No | 10 | 150 | 8.6 |
| **2M0355+1133** | … | … | 11.5 | L3 | … | 29+/-2 | 120+/-10 | 2017/1/13 | No | 14 | 120 | 117.7 |
| **KPNO Tau 4** | … | … | 13.3 | M9.5 | … | 25+/-2.5 | 2+/-1 | 2017/1/13 | No | 18 | 166 | 35.6 |

**Supplementary Table 2. Parameters Used to Generate Atmospheric Models and Best-Fit Rotation Rates, Barycentric Radial Velocity Offsets.**

| System | $T_{eff}$ (K) | log(g) | $vsini_{pl}$ (km/s) | $RV_{pl}$ (km/s) |
|---|---|---|---|---|
| **ROXs 42B** | 2100 | 3.81 | 9.5 (+2.1-2.3) | -2.3 +/- 4.0 |
| **GSC 6214-210** | 2188 | 4.05 | 6.1 (+4.9 -3.8) | -7.3 +/- 4.0 |
| **VHS 1256-1257** | 1280 | 4.5 | 13.5 (+3.6 -4.1) | 2.1 (+1.6 -1.7) |
| **OPH 90** | 2100 | 3.81 | 8.4 (+5.5 -5.0) | 7.8 (+1.3 -1.2) |
| **USco 1608-2315** | 2442 | 3.95 | 16.3 (+2.4 -2.5) | -4.0 (+1.1 -1.0) |
| **PSO J318.5-22** | 1325 | 3.7 | 12.0 (+3.5 -4.4) | -6.8+/-0.7 |
| **KPNO Tau 4** | 2477 | 3.74 | 16.3 (+3.2 -3.8) | 17.9 (+0.7 -0.8) |
| **2M0355+1133** | 1905 | 4.75 | 14.7 (+2.1 -2.3) | 11.8 +/- 0.5 |

For PSO J318.5-22 $T_{eff}$ and log(g) came from Allers et al 2016[38]. We note that the $T_{eff}$ determined by forward modeling the spectrum of PSO J318.5-22 is higher than that inferred from evolutionary models[22,38], suggesting that atmospheric models over-predict $T_{eff}$. Similarly, we adopt a higher $T_{eff}$ for our atmospheric model for VHS 1256-1257 b than would be inferred from COND models using its mass and age, since the lower temperature models predict a significant abundance of methane that is not seen in our spectrum. All other $T_{eff}$ and log(g) listed in this table come from COND models, where we selected temperature and surface gravity values that corresponded to masses and ages closest to inferred masses and ages of each object (Supplementary Table 1).

We determined the barycentric velocity correction for each of our RV measurements using the program barycorr[44]. We note that since we only had two AB pairs for the host star GSC 6214-210, we were not able to obtain accurate uncertainty estimates for rotation rate and RV from the MCMC analysis. We therefore adopt more robust uncertainties from our analysis of VHS 1256-1257. Rotation rates have previously been measured for 2M0355+1133[45], KPNO Tau 4[14], and PSO J318.5-22[38] with published values of 12.31+/-0.15 km/s, 10+/-2 km/s, and 17.5 (+2.3 -2.8) km/s, consistent with the measured rotation rates in this paper at 1.0σ, 1.3σ, and 1.2σ respectively. We note that the previously published spin measurements for KPNO Tau 4 and 2M0355+1133 used models including pressure broadening while our models did not, and we would expect the inclusion of pressure broadening to reduce the reported rotation rates by several km/s (see Methods section 3 for more details). RVs have previously been measured for 2M0355+1133 and PSO J318.5-22 with values of 11.92+/-0.22 km/s and -6.0 (+0.8 -1.1) km/s, consistent with our measured values at 0.2σ and 0.6σ. See Methods section 3 for a discussion of the reported RV uncertainties for companions ROXs 42B b and GSC 6214-210 b.

**Supplementary Table 3. Brown Dwarf Properties, Including New Homogenous Mass Estimates and Rotation Rates From the Literature.**

| System | RA | Dec | SpT | Kmag (mag) | Dist. (pc) | Age (Myr) | Mass ($M_{Jup}$) | +1σ ($M_{Jup}$) | -1σ ($M_{Jup}$) | *vsini*/$v_{eq}$ (km/s) | Ref. |
|---|---|---|---|---|---|---|---|---|---|---|---|
| **2M1139-3159** | 11 39 51.140 | -31 59 21.50 | M8 | 11.503 +/- 0.023 | 50+/-1.8 | 10+/-2 | 23.174 | 1.409 | 1.854 | 25.5 | *15, 16* |
| **2M1207-3932** | 12 07 33.500 | -39 32 54.40 | M8 | 11.945 +/- 0.026 | 50+/-1.8 | 10+/-2 | 19.916 | 0.925 | 1.187 | 13.7 | *15, 16* |
| **2M 05373648-0241567** | 05 37 36.480 | -02 41 56.70 | M7 | 14.560 +/- 0.100 | 442 +/- 20 | 2-3 | 46.262 | 5.485 | 5.288 | 38.18 | *46* |
| **CFHT-BD-Tau 1** | 04 34 15.272 | 22 50 30.96 | M7 | 11.849 +/- 0.018 | 145 +/- 15 | 2+/-1 | 54.128 | 6.362 | 10.249 | 7.0 | *15* |
| **CFHT-BD-Tau 2** | 04 36 10.387 | 22 59 56.03 | M7.5 | 12.169 +/- 0.019 | 145 +/- 15 | 2+/-1 | 43.240 | 4.477 | 7.502 | 8.0 | *15* |
| **CFHT-BD-Tau 3** | 04 36 38.938 | 22 58 11.90 | M7.75 | 12.367 +/- 0.025 | 145 +/- 15 | 2+/-1 | 38.652 | 3.544 | 5.619 | 12.0 | *15* |
| **Cha Ha 1** | 11 07 17.0 | -77 35 54.00 | M7.75 | 12.174 +/- 0.024 | 160 | 2 | 47.311 | 6.025 | 6.916 | 7.6 | *47* |
| **GG Tau Bb** | 04 32 30.25 | 17 31 30.90 | M7.5 | 12.010 +/- 0.130 | 145 +/- 15 | 2+/-1 | 46.940 | 7.189 | 7.991 | 6.8 | *15, 48* |
| **IC 348 355** | 03 44 29.210 | 32 08 13.70 | M8 | 13.499 +/- 0.035 | 300 | 1-3 | 47.717 | 4.030 | 5.540 | 45.0 | *15* |
| **IC 348 363** | 03 44 17.265 | 32 00 15.23 | M8 | 13.695 +/- 0.038 | 300 | 1-3 | 43.098 | 3.503 | 4.439 | 14.0 | *15* |
| **IC 348 405** | 03 44 21.163 | 32 06 16.56 | M8 | 13.910 +/- 0.100 | 300 | 1-3 | 38.659 | 3.186 | 3.820 | 45.0 | *15* |
| **KPNO Tau 5** | 04 29 45.680 | 26 30 46.81 | M7.5 | 11.536 +/- 0.018 | 145 +/- 15 | 2+/-1 | 60.490 | 17.359 | 16.611 | 10.0 | *15* |
| **USco 130** | 15 59 43.665 | -20 14 39.61 | M7 | 13.075 +/- 0.034 | 145 +/- 15 | 11+/-2 | 56.487 | 16.303 | 23.397 | 14.6 | *15, 16* |
| **USco 131** | 16 00 19.443 | -22 56 28.77 | M7 | 13.481 +/- 0.033 | 145 +/- 15 | 11+/-2 | 35.950 | 10.238 | 13.445 | 16.0 | *15* |
| **USco DENIS 161916** | 16 19 16.463 | -23 47 23.54 | M8 | 13.596 +/- 0.050 | 145 +/- 15 | 11+/-2 | 26.305 | 3.760 | 5.949 | 5.0 | *15* |
| **USco DENIS 162041** | 16 20 41.445 | -24 25 49.17 | M7.5 | 12.902 +/- 0.019 | 145 +/- 15 | 11+/-2 | 60.859 | 21.923 | 18.916 | 16.0 | *15* |
| **2M 05375206-0236046** | 05 37 52.060 | -02 36 04.60 | M6.5 | 14.200 +/- 0.060 | 442 +/- 20 | 2-3 | 59.575 | 8.426 | 13.313 | 14.8 | *46* |
| **2M05391308-0237509** | 05 39 13.080 | -02 37 50.90 | M7 | 14.310 +/- 0.070 | 442 +/- 20 | 2-3 | 60.490 | 17.359 | 11.971 | 15.4 | *46* |
| **2M05400453-0236421** | 05 40 04.530 | -02 36 42.10 | M6.5 | 14.270 +/- 0.070 | 442 +/- 20 | 2-3 | 57.297 | 5.469 | 8.372 | 40.3 | *46* |
| **Cha Ha 12** | 11 05 37.5 | -77 43 07.0 | M6.5 | 11.811 +/- 0.019 | 160 | 2 | 76.412 | 17.581 | 24.264 | 25.7 | *47* |
| **GY 37** | 16 26 27.810 | -24 26 41.82 | M6 | 12.092 +/- 0.030 | 120 +/- 10 | 3+/-2 | 45.420 | 6.335 | 10.583 | 22.5 | *15* |
| **IC 348 478** | 03 44 35.937 | 32 11 17.51 | M6.25 | 14.574 +/- 0.073 | 300 | 1-3 | 34.119 | 2.550 | 2.885 | 20.0 | *15* |
| **GY 141** | 16 26 51.284 | -24 32 41.99 | M8.5 | 13.889 +/- 0.057 | 120 +/- 10 | 3+/-2 | 17.953 | 1.877 | 1.961 | 4.9 | *15, 17* |
| **KPNO Tau 1** | 04 15 14.714 | 28 00 09.61 | M8.5 | 13.772 +/- 0.035 | 145 +/- 15 | 2+/-1 | 21.394 | 1.867 | 2.558 | 5.5 | *15* |
| **KPNO Tau 12** | 04 19 01.270 | 28 02 48.70 | M9 | 14.927 +/- 0.092 | 145 +/- 15 | 2+/-1 | 12.664 | 1.632 | 1.769 | 5.0 | *15* |
| **KPNO Tau 6** | 04 30 07.244 | 26 08 20.79 | M8.5 | 13.689 +/- 0.037 | 145 +/- 15 | 2+/-1 | 22.047 | 2.023 | 2.4733.032 | 5.0 | *15* |
| **S Ori 45** | 05 38 25.500 | -02 48 36.00 | M8.5 | 15.690 +/- 0.212 | 442 +/- 20 | 2-3 | 26.012 | 2.487 | 3.091 | 151 | *49, 50* |
| **TWA 5B** | 11 31 55.400 | -34 36 29.00 | M8.5 | 11.400 +/- 0.200 | 50+/-1.8 | 10+/-2 | 23.615 | 2.134 | 3.692 | 16.8 | *15, 16* |
| **USco DENIS 161006** | 16 10 06.082 | -21 27 44.02 | M8.5 | 13.768 +/- 0.056 | 145 +/- 15 | 11+/-2 | 23.328 | 2.153 | 2.244 | 7.0 | *15* |
| ***OPH 90*** | 16 27 36.59 | -24 51 36.1 | L0 | 14.85+/-0.05 | 120 +/- 10 | 3+/-2 | 11.243 | 1.545 | 1.670 | 8.4 | This paper |
| ***USco 1608*** | 16 08 27.47 | -23 15 10.4 | L1 | 14.205 +/- 0.070 | 145 +/- 15 | 11+/-2 | 19.157 | 1.226 | 2.128 | 16.4 | This paper |
| ***2M0355 +1133*** | 03 55 23.37 | 11 33 43.7 | L3 | 11.526 +- 0.021 | 9.1+/-0.1 | 120+/-10 | 28.667 | 2.618 | 2.794 | 14.7 | This paper |
| ***KPNO Tau 4*** | 04 27 28.0 | 26 12 04.7 | M9.5 | 13.281 +/- 0.032 | 145 +/- 15 | 2+/-1 | 24.970 | 2.794 | | 16.3 | This paper |

We create this list by first identifying all brown dwarfs in the literature with spectral types later than M6, well-constrained ages typically less than 20 Myr, and measured rotation rates or rotation periods. We then derive new mass estimates using published magnitudes, spectral types, distances, and ages. Here we list median masses as well as uncertainties corresponding to the highest prior density for objects with estimated masses less than 80 $M_{Jup}$. Objects in italics have new measured rotation rates from our NIRSPEC program. For objects 2M05373648-0241567,

2M05375206-0236046, 2M05391308-0237509, 2M05400453-0236421, and S Ori 45, the rotation rates presented here are equatorial rotation rates determined from published photometric rotation periods; all others are projected rotation rates from measurements of rotational line broadening. The "Ref." column cites the reference where we obtained the rotation rates for each object.

**Additional References**